\documentclass[twocolumn,tighten,times]{aastex62}

\newcommand{\swift}{\textit{Swift}}
\newcommand{\bexrb}{BeXRB}
\newcommand{\bexrbs}{BeXRBs}

\newcommand{\integral}{\textit{INTEGRAL}}  
\newcommand{\xmm}{\textit{XMM-Newton}}
\newcommand{\chandra}{\textit{Chandra}}
\newcommand{\rosat}{\textit{ROSAT}}
\newcommand{\fermi}{\textit{Fermi}}
\newcommand{\maxi}{\textit{MAXI}}
\newcommand{\rxte}{\textit{RXTE}}
\newcommand{\smcxthree}{SMC~X-3}
\newcommand{\nustar}{\textit{NuSTAR}}
\newcommand{\scubed}{S-CUBED}
\newcommand{\typeone}{Type~I}
\newcommand{\typetwo}{Type~II}
\newcommand{\onex}{1SCUBEDX}
\newcommand{\replacedref}[2]{\editpre{#2}} 
\newcommand{\editpre}[1]{#1}
\received{2018 March 21}
\revised{2018 October 11}
\accepted{2018 October 11}

\submitjournal{ApJS}

\shorttitle{\scubed{}}
\shortauthors{Kennea et al.}

\begin{document}

\title{The First Year of \scubed{}: The \swift{} Small Magellanic Cloud Survey}

\correspondingauthor{J.~A. Kennea}
\email{jak51@psu.edu}

\author[0000-0002-6745-4790]{J.~A. Kennea}
\affil{Department of Astronomy and Astrophysics, The Pennsylvania State University, University Park, PA 16802, USA}

\author[0000-0002-0763-8547]{M.~J. Coe}
\affil{Physics and Astronomy, University of Southampton, SO17 1BJ, UK}

\author[0000-0002-8465-3353]{P.~A. Evans}
\affil{University of Leicester, X-ray and Observational Astronomy Research Group, Leicester Institute for Space and Earth Observation, Department of Physics \& Astronomy, University Road, Leicester, LE1 7RH, UK}

\author{J. Waters}
\affil{Physics and Astronomy, University of Southampton, SO17 1BJ, UK}

\author{R.~E. Jasko}
\affil{Department of Astronomy and Astrophysics, The Pennsylvania State University, University Park, PA 16802, USA}

\begin{abstract}

The \swift{} Small Magellanic Cloud Survey, \scubed{}, is a high cadence shallow X-ray survey of the SMC. The survey consists of 142 tiled pointings covering the optical extent of the SMC, which is performed weekly by \replacedref{the NASA \swift{} Mission}{NASA's Neil Gehrels Swift Observatory}, with an exposure per tile of 60 seconds. The survey is focused on discovery and monitoring of X-ray outbursts from the large known and unknown population of \bexrbs{} in the SMC. Given the very low background of \swift{}'s X-ray telescope, even with a short exposure per tile, \scubed{} is typically sensitive to outbursts in the SMC at $>1-2\%$ Eddington Luminosity for a typical $1.4M_\odot$ neutron star compact object. This sensitivity, combined with the high cadence, and the fact that the survey can be performed all year round, make it a powerful discovery tool for outbursting accreting X-ray pulsars in the SMC. In this paper describe results from the first year of observations of \scubed{}, which includes \deleted{a} the \onex{} catalog of 265 X-ray sources, 160 of \editpre{which} are not identified with any previously cataloged X-ray source. We report on bulk properties \editpre{of} sources in the \onex{} catalog. Finally we focus on results of \scubed{} observations of several interesting sources, which includes discovery of three \typetwo{} outbursts from \bexrbs{}, and the detection of \typeone{} outbursts and orbital periods in 6 \deleted{other} \bexrb{} systems.

\end{abstract}


\section{Introduction}\label{sec:intro}

The Small Magellanic Cloud (SMC) is an irregular dwarf galaxy located at a distance of approximately $62$~kpc (e.g.~\citealt{Haschke12}). In X-rays the SMC is extremely active, with the majority of the X-ray sources being high mass X-ray binaries (HMXBs), see for example \citep{CoeKirk15}. Many of these HMXB systems are X-ray pulsars \citep{Yang17} and approximately 98\% are known to have Be-star companions \citep{Coe05}. 

Be/X-ray Binaries (\bexrbs{}) are High-Mass X-ray Binaries, typically containing a neutron star (NS) compact object, as evidenced by the detection of X-ray pulsations in many of these objects. \bexrbs{} with black hole compact objects have also been proposed, but so far only one system has been identified, MWC 656 \citep{Casares14}. In \bexrbs{} the companion star is a massive B-type star, which shows Balmer lines in emission, which leads to the Be-star classification. The emission lines arise from the presence of a circumstellar disk around the star, and it is the material in this disk that provides the fuel for accretion on to the compact object, and hence gives rise to its X-ray emission. 

\bexrbs{} show variability across a broad range of timescales and wavelengths. In X-ray, the variability of \bexrbs{} is often characterized as \typeone{} and \typetwo{} outbursts. \typeone{} outbursts occur regularly, and are thought to be caused by the periastron passage of the NS passing close to the outer edge of the circumstellar disk. The regularity of \typeone{} bursts is therefore linked to the orbital period of the system. \typetwo{} outbursts are much brighter, getting close to Eddington, and sometimes super-Eddington, luminosities \citep{Townsend2017}, these outbursts can last for several orbital periods and arise when the mass ejection from the Be-star promotes an exceptionally large disk, often filling the entire orbit of the compact object. 

The SMC has an over-abundance of known \bexrb\ systems compared to the Milky Way due to recent high periods of star formation \citep{Harris04}. It also benefits observationally from a low foreground extinction ($N_\mathrm{H} = 5.34 \times 10^{20}\ \mathrm{cm}^{-2}$; \citealt{Willingale13}), and SMC X-ray sources have a relatively well constrained distance compared to those in the Milky Way. All of these factors make estimating the X-ray luminosity more accurate than for Milky Way objects. Consequently, the SMC provides an ideal laboratory for the discovery and study of \bexrbs{} in outburst. 

Previous observations of the SMC have been mostly focused on deep observations of the SMC, \editpre{e.g.~those surveys by} \rosat\ \citep{Kahabka96,Haberl00}, \xmm\ \citep{Sturm13} and \chandra\ \citep{Hong17,Antoniou09,McGowan08,Schurch07}. However, an irregular, approximately weekly, survey of selected regions of the SMC was performed over a period of more than a decade \citep{Galache08} by the Proportional Counter Array (PCA) on the Rossi X-ray Timing Explorer (\rxte; \citealt{Bradt93}). \rxte{} was deactivated in 2012 January, ending those observations. Without regular sensitive monitoring observations of the SMC, discovery of transient outbursts was limited to either those transients bright enough to be detected by all-sky survey instruments such as \swift's Burst Alert Telescope (BAT; \citealt{Barthelmy05}), \fermi{}'s Gamma-Ray Burst Monitor (GBM; \citealt{Meegan09}) and the ``Monitor of the All-sky X-ray Image'' (\maxi; \citealt{Matsuoka09}), or through infrequent scans of the SMC performed by \integral{} \citep{Coe10}. In all these cases sensitivities of the telescopes involved were such that they were only capable of detecting the few extremely bright \typetwo{} outbursts.

NASA's Neil Gehrels Swift Observatory (\swift{}; \citealt{Gehrels04}) is medium-sized Explorer (MIDEX) class satellite, launched in November of 2004, with the primary goal to study gamma-ray bursts (GRBs). \swift{} consists of three co-aligned instruments: BAT, which operates in the 15-150 keV energy range with a 1.4~sr field of view (FOV); the X-ray Telescope (XRT; \citealt{Burrows05}) with a 23.6 arcminute FOV operating in the 0.3--10~keV band; and the Ultra-violet/Optical Telescope (UVOT; \citealt{Roming05}), with a 17 arcminute field of view, observing at wavelengths between 170--650 nm.

The \swift{} SMC Survey (hereafter \scubed{}) was designed to harness the unique capabilities of \swift{}: its rapid slewing, which allows for low overhead observing with very short exposure times; and its sensitive low-background XRT, to perform weekly X-ray observations of the SMC, in order to both hunt for outbursting \bexrb{} sources, and to monitor the flux of transient and persistent sources. 

Although the SMC is relatively compact, the XRT's FOV only covers $\sim 0.12\ \mathrm{deg}^2$, therefore to cover a significant fraction of the SMC \replacedref{would require}{requires} many pointings. To achieve this we utilized a new short exposure tiling mode, developed to enable \swift{} to search large area (10s to 100s of square degrees) error regions associated with Gravitational Wave detections by the Advanced LIGO and Advanced Virgo detectors (e.g.~\citealt{Evans16}). Although this observation mode requires that each exposure is short (60~s) to cover this large area, \scubed{} is sensitive to outbursts from SMC sources at a level of $>3.5\times10^{36}\ \mathrm{erg\ s^{-1}}$, or $>2\%$ Eddington, for a typical \bexrb{} spectrum.

In this paper we describe the design, implementation, and results from \scubed{}, covering the first year of observations, which for the purposes of this paper includes all \scubed{} observations which were taken between 2016 June 8 and 2017 June 6. In addition, where follow-up Target of Opportunity (TOO) observations triggered by \scubed{} were taken by \swift{} or other observatories, we report on results of those observations, \editpre{if not reported elsewhere}. 

This paper is novel in three ways. Firstly, it describes a new method of performing X-ray surveys of large regions of the sky. \deleted{that has not previously been possible} \scubed{} represents the test bed \replacedref{for}{of} a \replacedref{new mode of observation by \swift{}}{new \swift{} tiled survey observing mode}, which would not have been possible with other \editpre{focused} X-ray telescopes, primarily as it is powered by \swift's fast slewing capability, which allows for low overhead \editpre{short} observations. Although regular scans of large regions such as the SMC have been performed by survey telescopes, or in scanning observations by missions such as \rxte{}, these surveys do not have either the high sensitivity that \scubed{} has, nor do they have sub-arcminute level spatial resolution required to accurately determine the source in outburst by localization.

Secondly, we report for the first time, the detection of large \typetwo{} outbursts of \bexrbs{} in the SMC, SXP~6.85 and SXP~59.0. In addition \scubed{} observations were the first to identify the major super-Eddington outburst of \smcxthree{}, which has been extensively reported on elsewhere (e.g.~\citealt{Townsend2017,Tsygankov17,Weng17,Koliopanos18}), and we present here \scubed{} focused results. These detections, along with detections of \typeone{} outbursts, show the power of the \scubed{} observing technique in monitoring the large population of \bexrbs{} in the SMC, including outbursts which would be below the typical sensitivity of larger area X-ray survey telescopes.

Thirdly, we present a catalog of all high-quality detections of X-ray point sources \deleted{detected} in the SMC. Although the \scubed{} combined first year data have a relatively shallow exposure time and lower sensitivity compared to previous X-ray surveys, such those performed by \xmm{}, \chandra{}, and \rosat{}, it covers both a larger spatial area \editpre{and longer time period} than those surveys, leading to the detection of 160 previously unknown X-ray sources, \deleted{detected with high confidence} the details of which are reported for the first time in this paper.

\section{Survey Design} \label{sec:survey}

The goals of \scubed{} are three-fold. Firstly, to provide full coverage of the known X-ray sources that are located in the optical extent of the SMC, including the region known as the ``Wing''. The decision to limit to the optical core of the SMC was made in order to limit the exposure time required to perform the survey, but also to avoid strongly biasing the survey towards observing regions that have been previously well studied in X-rays. Secondly, to perform the search at a sufficient level of sensitivity so that a transient can be detected \replacedref{early to enable triggering on outbursting sources that normally would not be noticed until they had reached near peak brightness by}{earlier than previously possible with} less sensitive all-sky monitor telescopes such as BAT and \maxi{}. Thirdly, to perform observations at a cadence sufficient for rapid reporting and good sampling of light-curves of typical \typetwo{} outbursts of \bexrbs. All of these goals need to be met in a way which is acceptable to the operation of \swift{}, and must not overburden the \swift{} \replacedref{science plan}{observing schedule} to the detriment of other observing programs.

In order to determine the area to tile, we utilized \deleted{with} the X-ray source catalogs of \rosat{} \citep{Haberl00} and \xmm{} \citep{Sturm13}. The tiling pattern was generated algorithmically, by searching over a region $6\degr \times 6\degr$ \editpre{centered on the SMC}, and ensuring a tile was placed within any region within two XRT FOV radii (23\farcm6\deleted {radius}) of regions where the source density was greater than 3 sources per field. 
No attempt \deleted{is made} to cover the LMC-SMC bridge was made, as this region covers too large an area. 

XRT has an approximately circular FOV with an 11.8~arcminute radius. \swift{}'s slewing is optimized for speed, which leads to a pointing error up to 3~arcminutes (i.e. \swift{} can land up to 3~arcminutes off target). To compensate for this, the overlap between tiles was set to 3~arcminutes, as this ensures gap-less coverage. The survey is purposely conservative in its approach to covering the SMC, in order to ensure that we are not biasing the survey to just observing previously known sources, but also those in the gaps between them, therefore the tiling is continuous.

\begin{figure}[t]
\centering
\includegraphics[width=0.5\textwidth]{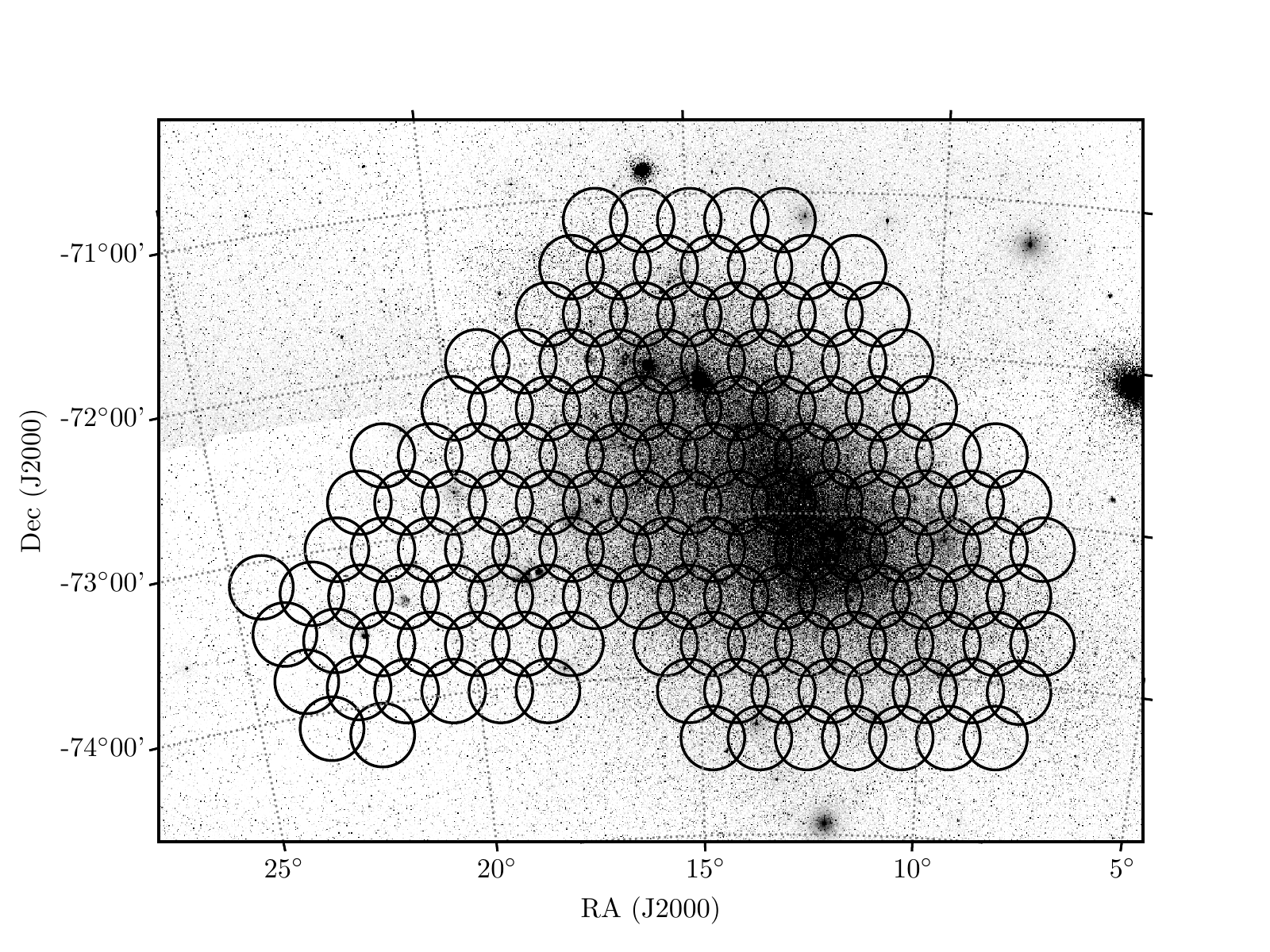}
\caption{\label{fig:survey}Configuration of the \scubed{} survey. Survey consists of 142 tiles covering the central optical extent of the SMC and Wing. Tiles overlaid over image from the Digitized Sky Survey.}
\end{figure}

The resulting configuration of the survey is shown in Figure~\ref{fig:survey}. \replacedref{The final survey}{\scubed{}} consists of 142 \swift{} pointings, covering an area of approximately 13.2 square degrees.

The XRT background is exceptionally low, the on-orbit instrumental background of \swift{} has been measured to be $10^{-6}\ \mathrm{counts}\ \mathrm{s}^{-1}\ \mathrm{pixel}^{-1}$ \citep{Evans14} and XRT's pixel scale is 2.36 arcseconds/pixel. This means that in an typical circular XRT source extraction region with a 20-pixel radius, the background would \editpre{be} $\sim1$ count in 1~ks. Assuming the minimum required number of counts required for an XRT detection is 5 counts, and typical spectral parameters for a \bexrb{} in the SMC, for a 60~s exposure we would detect any source brighter than a luminosity of
2\% of $L_\mathrm{Edd}$.
For a softer spectrum source (for example a Crab like spectrum), this sensitivity is increased to 1\% of $L_\mathrm{Edd}$. Given the typical outburst brightness of \bexrbs{}, which can reach super-Eddington levels (e.g.~\smcxthree{}; \citealt{Townsend2017}), it was determined that utilizing the minimum exposure time allowed would give sufficient sensitivity, as well as minimizing the load on the \swift{} schedule.

For cadence, typical orbital periods of \bexrbs{} are of the order of 10-100~days, and outburst rise times typically scale to be approximately on the order of 1 orbital period. Due to limitations of on-board memory, normal tiling can only be performed when single-day length plans are uploaded to \swift{} (Tuesday to Thursday), so the \swift{} Flight Operations Team preferred that the \scubed{} survey be scheduled at intervals of a fixed number of weeks, to ensure that the observations occurred on the same day of the week. Therefore in order to maximize our coverage of outbursts, and to ensure the quickest response time to new outbursting sources, it was decided to schedule \scubed{} at a cadence of once per week (see next section).

\section{Observations} \label{sec:obs}

\scubed{} observations were performed \deleted{weekly, }with the aim of observing \deleted{at least} 142 tiles covering the SMC every week, with an individual exposure per tile of 60~s, \replacedref{for}{to} a total of 8.52~ks \deleted{exposure per week}, not including slewing overheads. Due to the short distance between the tiles in the survey, the median slewing time between tiles was 23 seconds. The \swift{} scheduling system cannot handle observations shorter than 5~min, and schedules targets at a 60~s time resolution, which would be inefficient for for scheduling 60~s exposure tiles. Therefore, a custom planning solution was developed where \scubed{} tiles are scheduled as a single pointing near the center of the survey, and then replaced with tiles, using high accuracy estimates for slewing and visibility.

Due to observing constraints, higher priority \deleted{observations} targets, or interruption by TOO and GRB observations, \scubed{} was not always observed to 100\% completion every week. \swift's orbit pole constraint, which typically lasts up to 10 days, was the primary cause of \scubed{} observations not being scheduled. 

\begin{figure}[t]
\centering
\includegraphics[width=0.5\textwidth]{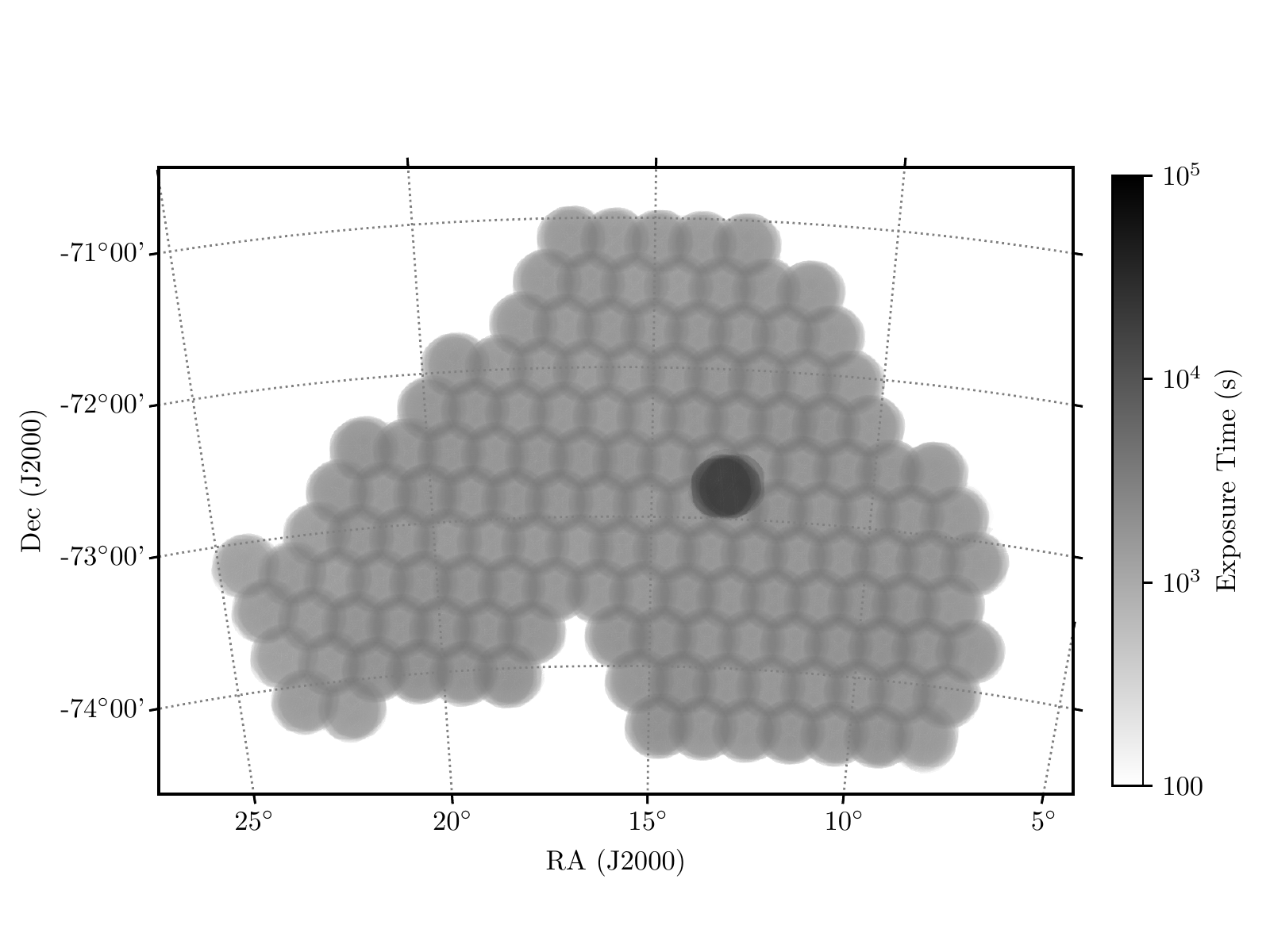}
\caption{\label{fig:expmap}Exposure map of the \scubed{} survey for the first year of observations. Individual TOO observation of \scubed{} triggered targets are not included in this map, as \scubed{} data only processes PC mode data, and TOOs are typically in WT to allow accurate measurement of the pulsar period. Note that the region of higher exposure is due to over-scheduling of observations, which means that after all tiles are scheduled, some time is spent observing a field at the approximate center of the SMC to fill the remaining time. } 
\end{figure}

In the first year, \scubed{} observations were performed \replacedref{for}{on} 43 out of 52 weeks, and the average completion rate for weeks when \scubed{} tiling was performed was 95\% of all tiles. Details of the individual tiling observations performed for \scubed{} are given in Table~\ref{tab:obs}. Note that observations in the first 5 weeks of \scubed{} were attempted at a cadence of every 8 days, in order to reduce UVOT filter wheel rotations by picking days on which the filter of the day was \textit{uvw1}, however, due to the limitation that the survey could not be performed on Friday$-$Monday, this was changed to a 7 day cadence, in order to perform the survey on Tuesdays. In some cases where the \scubed{} survey was not completed in a single day, observations were performed on the following days in order to make up lost time. An image of the exposure map of \scubed{} is shown in Figure~\ref{fig:expmap}. The median exposure of a non-overlapping region of the \scubed{} survey was 1919~s.

As it is often hard to predict the amount of slew-time required between tiles, sometimes more time was scheduled than was required to observe the 142 \scubed{} tiles. In those cases additional time was spent on a field which centered on the SMC, which can be clearly seen in Figure~\ref{fig:expmap} as a region of enhanced exposure. 

When, as the result of \scubed{} observations, an outburst of a X-ray transient source in the SMC is found, follow-up additional observations were requested through TOO requests to \swift{} or other telescopes such as \nustar{} and \chandra{} as appropriate, in order to obtain higher quality spectral and timing information. We report on the results from those observations along with \scubed{} data in Section~\ref{sec:results}. 

\startlongtable
\begin{deluxetable*}{llllrr}
 \tablecolumns{5}
\tablecaption{\label{tab:obs}Observations performed by \swift{} for the \scubed{} program in the first year of the program. In total \scubed{} tilings were performed 43 during the first year of observations, which started on 2016 June 8 and ended 2017 June 6. Missing weeks are typically caused by the SMC being pole constrained (see text), or by the observations being not scheduled due to higher priority observations.}
\tablewidth{0pt}
\tablehead{\colhead{Week} & \colhead{Start time (UTC)} & \colhead{End Time (UTC)} & \colhead{Tiles Observed} & \colhead{Exposure}}
\startdata
1 & 2016 June 08 00:06:02 & 2016 June 08 00:06:57 & 142 (100\%) & 8.5 ks &  \\
2 & 2016 June 16 00:06:02 & 2016 June 16 01:06:11 & 78 (55\%) & 4.6 ks &  \\
3$\dag$ & 2016 June 24 00:06:02 & 2016 June 24 00:06:52 & 127 (89\%) & 7.5 ks &  \\
4$\dag$ & 2016 June 28 01:06:02 & 2016 June 28 01:06:34 & 142 (100\%) & 8.9 ks &  \\
5$\dag$ & 2016 July 06 02:07:01 & 2016 July 06 02:07:52 & 142 (100\%) & 8.8 ks &  \\
5$\dag$ & 2016 July 10 08:07:02 & 2016 July 10 08:07:59 & 142 (100\%) & 8.7 ks &  \\
6 & 2016 July 15 06:07:02 & 2016 July 15 06:07:55 & 142 (100\%) & 8.8 ks &  \\
8 & 2016 July 29 04:07:02 & 2016 July 29 04:07:42 & 139 (98\%) & 8.6 ks &  \\
9 & 2016 August 03 00:08:02 & 2016 August 03 00:08:28 & 139 (98\%) & 8.9 ks &  \\
10 & 2016 August 10 00:08:02 & 2016 August 10 00:08:34 & 142 (100\%) & 8.9 ks &  \\
11 & 2016 August 17 01:08:58 & 2016 August 17 01:08:06 & 138 (97\%) & 8.2 ks &  \\
12 & 2016 August 24 06:08:02 & 2016 August 24 06:08:39 & 142 (100\%) & 8.8 ks &  \\
13 & 2016 August 31 01:08:02 & 2016 August 31 01:08:32 & 142 (100\%) & 8.9 ks &  \\
16 & 2016 September 21 00:09:02 & 2016 September 21 00:09:02 & 142 (100\%) & 8.9 ks &  \\
17 & 2016 September 28 04:09:02 & 2016 September 28 04:09:37 & 99 (70\%) & 6.0 ks &  \\
18 & 2016 October 05 08:10:02 & 2016 October 05 08:10:27 & 142 (100\%) & 8.7 ks &  \\
19 & 2016 October 12 05:10:02 & 2016 October 12 06:10:51 & 142 (100\%) & 8.6 ks &  \\
20 & 2016 October 19 07:10:02 & 2016 October 19 07:10:31 & 137 (96\%) & 8.5 ks &  \\
21 & 2016 October 25 06:10:02 & 2016 October 25 06:10:28 & 145 (100\%) & 9.3 ks &  \\
23 & 2016 November 09 09:11:02 & 2016 November 09 09:11:03 & 142 (100\%) & 8.7 ks &  \\
24 & 2016 November 16 00:11:02 & 2016 November 16 00:11:34 & 142 (100\%) & 8.7 ks &  \\
25 & 2016 November 23 13:11:02 & 2016 November 23 13:11:33 & 139 (98\%) & 8.3 ks &  \\
26 & 2016 November 30 00:11:02 & 2016 November 30 00:11:21 & 142 (100\%) & 8.5 ks &  \\
27 & 2016 December 07 04:12:02 & 2016 December 07 04:12:09 & 142 (100\%) & 8.9 ks &  \\
28 & 2016 December 14 00:12:02 & 2016 December 14 00:12:57 & 140 (99\%) & 8.7 ks &  \\
30 & 2016 December 28 06:12:02 & 2016 December 28 06:12:56 & 142 (100\%) & 8.5 ks &  \\
31 & 2017 January 05 12:01:02 & 2017 January 05 12:01:01 & 142 (100\%) & 8.8 ks &  \\
32 & 2017 January 11 02:01:02 & 2017 January 11 02:01:33 & 142 (100\%) & 9.5 ks &  \\
33 & 2017 January 18 01:01:02 & 2017 January 18 01:01:10 & 107 (75\%) & 6.5 ks &  \\
34 & 2017 January 25 00:01:02 & 2017 January 25 00:01:12 & 142 (100\%) & 8.5 ks &  \\
35 & 2017 February 01 00:02:02 & 2017 February 01 00:02:31 & 142 (100\%) & 8.7 ks &  \\
38 & 2017 February 22 00:02:02 & 2017 February 22 00:02:02 & 142 (100\%) & 8.4 ks &  \\
40 & 2017 March 08 00:03:02 & 2017 March 08 00:03:57 & 122 (86\%) & 7.3 ks &  \\
41 & 2017 March 14 02:03:02 & 2017 March 14 02:03:26 & 140 (99\%) & 8.6 ks &  \\
42 & 2017 March 22 00:03:02 & 2017 March 22 00:03:31 & 117 (82\%) & 7.2 ks &  \\
43 & 2017 March 29 12:03:02 & 2017 March 29 13:03:20 & 136 (96\%) & 8.7 ks &  \\
45 & 2017 April 11 01:04:02 & 2017 April 11 01:04:37 & 126 (89\%) & 8.1 ks &  \\
46 & 2017 April 18 00:04:02 & 2017 April 18 00:04:07 & 142 (100\%) & 8.7 ks &  \\
47 & 2017 April 25 00:04:02 & 2017 April 25 00:04:38 & 116 (82\%) & 6.9 ks &  \\
48 & 2017 May 02 00:05:02 & 2017 May 02 00:05:23 & 140 (99\%) & 8.6 ks &  \\
49 & 2017 May 09 00:05:02 & 2017 May 09 00:05:32 & 140 (99\%) & 8.5 ks &  \\
50 & 2017 May 16 01:05:02 & 2017 May 16 01:05:38 & 129 (91\%) & 8.2 ks &  \\
53 & 2017 June 06 00:06:02 & 2017 June 06 00:06:05 & 117 (82\%) & 7.2 ks &  \\
\enddata
\tablenotetext{\dag}{Note that in weeks 3 and 4 and week 5, \scubed{} observations were taken at a slightly higher cadence (every 4 days). This resulted in there being two sets of tiling observations performed in week 5, hence why week 5 appears twice in the table.}.
\end{deluxetable*}

\section{Data Analysis}\label{sec:analysis}

Analysis of large tiled regions is computationally intensive, and combining the entire \scubed{} \editpre{surveyed region} proved to be too slow and memory intensive. However, analysis of individual tiles creates a different problem, how to deal with the fact that tiles overlap and many sources will appear in more than one tile. To achieve maximum sensitivity, it is best to perform some combining of tiles, but the number of tiles cannot be so large as to make the analysis computationally slow, as we are aiming for near-real time reporting of transients. 

Therefore the individual XRT tiles were grouped into `blocks', which includes all tiles up to $\sim0.6\degr$ in radius from the center of the block (corresponding to three XRT fields in diameter). Blocks were defined such that every field and every overlap between fields was included in at least one block, while creating the minimum number of blocks necessary to ensure this. Analysis was then performed on a block-by-block basis as described below. Note that some fields are in multiple blocks, in which case the receipt of data triggered the analysis of multiple blocks; such analyses were performed independently.

To analyze a block, source detection was carried out on using the iterative cell-detect algorithm of \cite{Evans14}, after every observation. This algorithm assigns each source a quality flag of `Good', `Reasonable' or `Poor' which relates to how likely it is to be a real astrophysical source. These adjectival ratings are defined so that for `Good' sources, we expect a false-positive rate of $0.3\%$, for 'Reasonable' a false positive rate of $1\%$, and for 'Poor' sources a false positive rate of $10\%$. These false positives rates were calculated utilizing extensive simulations as described by \citealt{Evans14} for the 1SXPS catalog, with the strict definitions of those flags listed in Table~11 of that paper. As the pipeline for detection used in \scubed{} is derived from that used by the 1SXPS catalog, the false-positive rate will be identical to those in calculated for 1SXPS.

This analysis was performed on the sum of all data collected for that block at the time of analysis, i.e.\ we did not search for sources in each weekly 60-s data set, but only in the summed data set of all \scubed{} observations taken so far. Therefore as the \scubed{} cumulative exposure built up, the analysis system is able to detect sources that are fainter, building up a large catalog of X-ray sources. All objects detected in analysis of each block were compared with the list of sources already detected in the \scubed{} survey; sources whose position agreed to within 5-$\sigma$ were identified with that source. If a source does not match a previously detected one, then it is added to the source list. Such new sources can potentially either be the result of the source brightening (i.e. a new transient), or simply the detection of a faint source made possible by the longer exposure.

This step of merging sources could only be called by one block at a time to prevent a newly-detectable source being added to the source list multiple times. For sources that are detected multiple times during the survey, the best position of the source was defined as that detection which yielded the best quality flag (i.e.\ the position from a `Good' detection supersedes a `Reasonable' detection). If multiple detections had the same quality flag, then the position was taken from the detection with the highest signal-to-noise ratio (S/N), and if there were multiple detections with the same quality flag and S/N, then that with the smallest position uncertainty was taken.

A list of all sources detected in the block, along with any pre-existing sources which lay within the block but were not detected in the latest analysis, was then produced. These sources were then each analyzed individually. The pipeline keeps track of which data were used in each source analysis, therefore if an observation of a given source was part of two blocks, the source analysis was only conducted once. 

For each source we produced a light curve, spectrum and, where possible, an ``enhanced'' position (utilizing UVOT to reduce the systematic error on astrometry), using the tools described by \cite{Goad07} and \cite{Evans07,Evans09}, which corrected for vignetting, dead columns on the CCD, pile up and other effects. Note that these tools are designed and calibrated for point sources only: the fluxes for extended sources such as the supernova remnant 1E~0102.2$-$7219 are not reliable. By default light curves were produced to have one bin per observation to give an at-a-glance idea of variability. However, for sources fainter than $\sim0.1\ \mathrm{count s^{-1}}$ such a light curve will be comprised entirely of upper limits because although the source was solidly detected in the summed data set, no single 60-s observation is enough to yield a flux measurement which is non-zero at the 3-$\sigma$ level. The automated analysis was made available to the \scubed{} team via a web interface, which allows easy re-binning of the light curve, allowing us to manually investigate the variability of such faint sources.

Due to the large number of sources detected it is impractical to manually examine each source for signs of outbursting activity. Instead the analysis software automatically identifies sources which may be transient or in outburst. For known sources, this is done by comparing the peak flux in the \scubed{} light curve with the cataloged flux; for previously unknown sources the peak \scubed{} flux is  compared with a 3-$\sigma$ upper limit calculated from the \replacedref{RASS}{\rosat{} All-Sky Survey \citep{Voges99}} at the location of the source. In either case, if the \scubed{} flux is at least 5 $\sigma$ above the comparison flux, the source is flagged.

For any source thus flagged, the pipeline sends a notification to the \scubed{} team members (keeping track of such notifications to avoid repeated alerts for a single outburst), which then allows for the light-curve to be checked by hand. Since this check is performed on the per-observation light curve, it implicitly requires a source to be bright enough to produce a detection in the light curve ($\sim0.1$ ct s$^{-1}$) before it can be classed as in outburst, thus it is possible that there are objects in our sample that have undergone outbursts at a lower level which are not identified by the real-time analysis software. Therefore for this paper we reanalyzed all pipeline generated light-curves to look for signs of transient behavior in them.

As described in Section~\ref{sec:obs}, in some cases we obtained additional observations of such sources with \swift{} to monitor the outburst. Such observations were not included in the automated analysis or when performing source detection; in this case source products were built using the on-line tools at \texttt{http://www.swift.ac.uk/user\_objects}, described by \cite{Evans09}.

For results reported in this paper we assume a mean SMC distance of 62~kpc, and unless otherwise stated all errors are $1\sigma$, except for localization errors which are given as a $90\%$ confidence radius.

\begin{figure*}[t]
    \centering
    \includegraphics[width=1.0\textwidth]{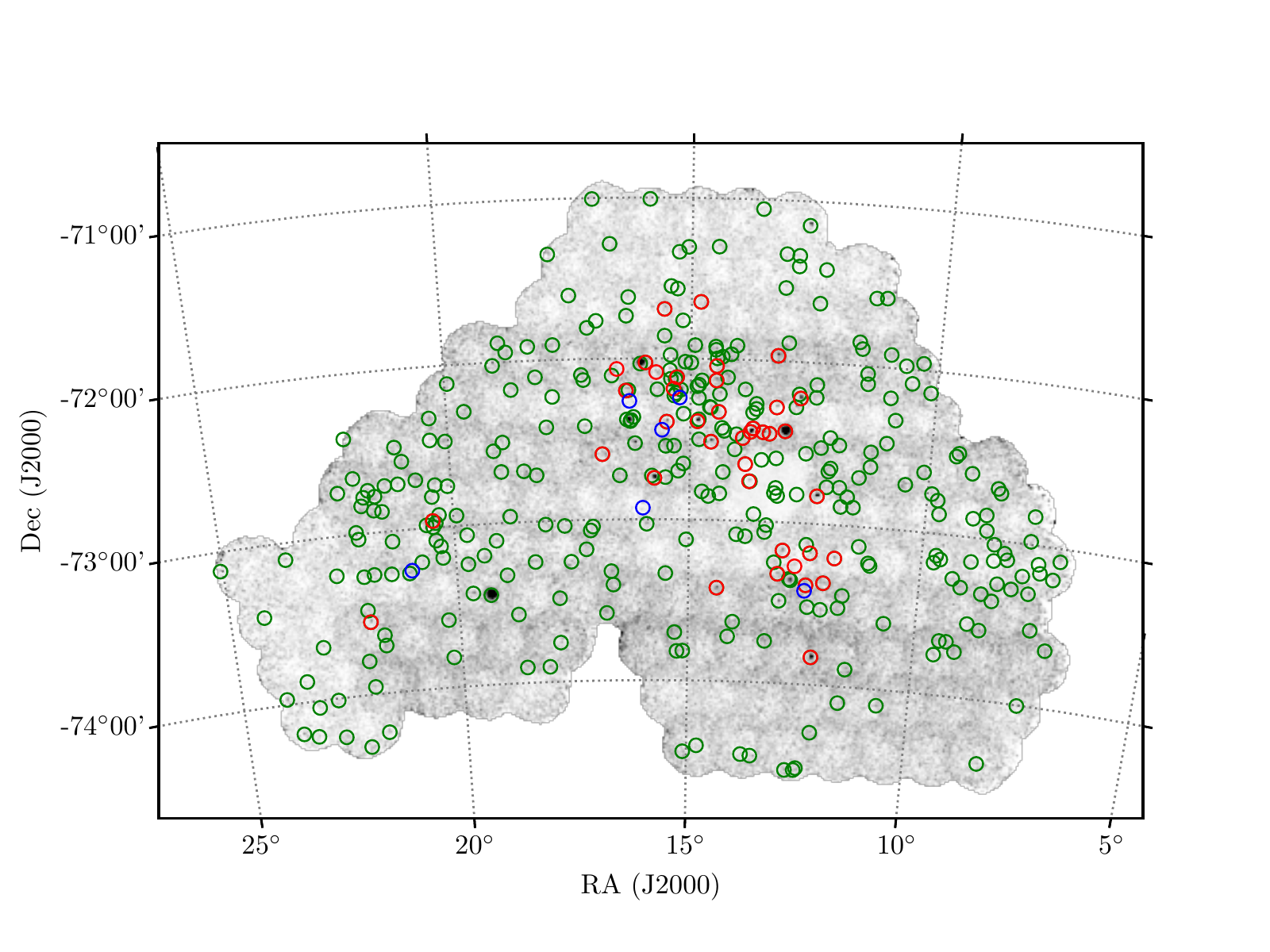}
    \caption{\label{fig:xrayimage}Combined exposure corrected \scubed{} X-ray map of the SMC. Note that variations in the background level are varying exposure times across the field, rather than structure in the actual emission. Lighter areas have longer exposure times, and therefore lower noise. Marked on this image are all sources flagged as `Good' by \scubed{}: Red circles are known ``SXP'' targets; blue are HMXBs identified in Table~\ref{tab:new}; green are all others.}
    \end{figure*}
    
\section{\onex{} - The First Year \scubed{} source catalog}

The combined \scubed{} X-ray image of the SMC, corrected for variations in exposure is shown in Figure~\ref{fig:xrayimage}. The median exposure time over the whole mosaic is $1919$~s, although the survey contains regions of higher exposure due to overlapping tiles and other over-exposed regions as shown in Figure~\ref{fig:expmap}. Prominent in the field are several bright well known X-ray sources, the brightest three of which are \smcxthree, which underwent an outburst in 2016-2017 \citep{Townsend2017}, SMC X-1, a persistent but highly variable HXMB (e.g.~\citealt{Li97}), and the X-ray bright supernova remnant 1E~0102.2-7219, which is most often utilized as an X-ray calibration source \citep{2017A&A...597A..35P}. A large number of other point sources are visible in the image. Absent is any strong diffuse X-ray emission, as expected due to the relative insensitivity of the \scubed{} survey. The primary purpose of \scubed{} is to examine variability of the X-ray point source population in the SMC, and to detect any turn-on of transient sources. However, in addition to this, \scubed{} also detected a great number of X-ray point sources that were previously unknown. In the this section we present the results of detection, localization, and \replacedref{flux monitoring}{periodicity searches} of point sources detected by \scubed, over the first year of observations, and present \editpre{the first year} catalog of \scubed{} sources: \editpre{\onex{}}.

Automated data analysis of the first year of \scubed{} observations as described in Section \ref{sec:analysis} detected a total of 808 point sources. Of these, 265 were flagged as `Good', and are therefore considered to have a high likelihood of being real X-ray point sources, 110 were `Reasonable' and 431 were `Poor'. To limit the scope of the results presented in this paper, we only report on those sources are flagged as `Good' by our data analysis. Based on the statistical simulations performed by \cite{Evans14}, the \deleted{expected} number of spurious sources flagged as `Good' is expected to be less than 1.

We present in Table~\ref{tab:cattable} a catalog of all the `Good' sources detected in X-ray during the first year of \scubed{} observations. Hereafter we refer to this catalog as \onex{}, and sources in this catalog are named with the convention \onex{}~J\textit{HHMMSS.s} $\pm$ \textit{DDMMSS}, based on their \scubed{} derived coordinates. In addition, as a shorthand for sources, we \deleted{also} refer to targets by an internal catalog number, of the form ``SC$n$'' where $n$ \deleted{is the internal \scubed{} catalog number, which} corresponds to the order of discovery by the \scubed{} analysis software. Hence the brightest source in the SMC, SMC~X-1 is SC1, the bright supernova remnant 1E~0102.2$-$7219 is SC2 and so on.

For sources where a spectral fit was possible, we quote a fitted photon index, assuming a standard SMC absorption.\deleted{ of $6\times10^{20}\ \mathrm{cm^{-2}}$.}
In addition we calculate from this fit a counts-to-flux ratio, and use this value to convert count rates in individual observations into fluxes. In case where a spectral fit is not possible, we calculate the mean flux by taking the mean count rate and multiplying it by the median counts-to-flux ratio, derived from the average spectra of all point sources with a spectral fit, of $3\times10^{-11}\ \mathrm{erg\ cm^{-2}\ count^{-1}}$.  The mean flux across all observations is given for each source.\deleted{, in cases where a spectral to the combined data was possible, we utilize this value. }

In addition, for each source we give a detection percentage, which indicates how often the source was detected in individual \scubed{} tiling observations. In cases where this value was zero, this indicates that the source was not detected in any individual observation, but was detected in the combined first year of observations.

Source positions were matched against X-ray source catalogs including the 1SXPS \citep{Evans14}, 3XMM-DR5 \citep{XMM-DR5} and HEASARC X-ray master catalogs\footnote{https://heasarc.gsfc.nasa.gov/W3Browse/all/xray.html}, using positional coincidence. \onex{} sources are classified into two categories: known (K) and unknown (U), which specifically refers to whether they are \deleted{being} previously identified X-ray emitters. In the \onex{} catalog, 105 sources are identified as known, and 160 are  unknown. Therefore \onex{} represents a significant increase in the number of \deleted{known} X-ray sources in the vicinity of the SMC, despite the relatively shallow overall exposure. In Table~\ref{tab:cattable}, for all known X-ray sources, we give a common catalog name for the X-ray source.

For X-ray sources where \deleted{a} the identification of the X-ray source is known, for example known \bexrbs{} and HMXBs as given by the catalogs of \cite{CoeKirk15} and \cite{Haberl16}, we list the common name and source type. In addition we have cross referenced the \onex{} catalog against several other catalogs of SMC sources, and where a positive match is made based on positional coincidence, the name of the source and the reported source type is given, along with references to which catalogs this source appears in. Based on this positional matching, we find 16 \onex{} sources consistent with known active galactic nuclei (AGN), 4 of which are detected in X-ray for the first time by \scubed{}.

Full investigation of source types for all \scubed{} X-ray sources is outside the scope of this paper, and will be presented in a future work.

\subsection{\onex{} completeness}

For any catalog, it is important to understand the expected completeness of the survey as a function of flux. As \onex{} was generated utilizing the same analysis methods as the 1SXPS catalog, we used the same calculations for \replacedref{estimated}{estimating} completeness that were formulated for that catalog. \cite{Evans14} describe in detail the simulation process utilized to obtain this completeness estimate, and Figure~14 within that paper shows the expected 50\% and 90\% completeness levels for sources classified as `Good' at a variety of exposures and flux levels. The median exposure time for the \onex{} catalog is 1919s. Based on the calculations for 1SXPS, the expected flux at which \replacedref{the median an exposure at the median 1919~s exposure time}{the median exposure} is $50\%$ complete, is $2\times10^{-13}\ \mathrm{erg\ cm^{-2}\ s^{-1}}$, and it will be $90\%$ complete above $4\times10^{-13}\ \mathrm{erg\ cm^{-2}\ s^{-1}}$.

However, as the exposure time of \scubed{} is not uniform, simply using the median exposure time as an estimate of completeness would underestimate the occurrence of faint sources in regions of higher exposure (e.g.~tile overlaps). Therefore, in order to better calculate the completeness of the entire \onex{} catalog, we performed a Monte-Carlo simulation based upon the configuration of \scubed{} observations taken in the first year. 

\begin{figure}[t]
\centering
\includegraphics[width=0.5\textwidth]{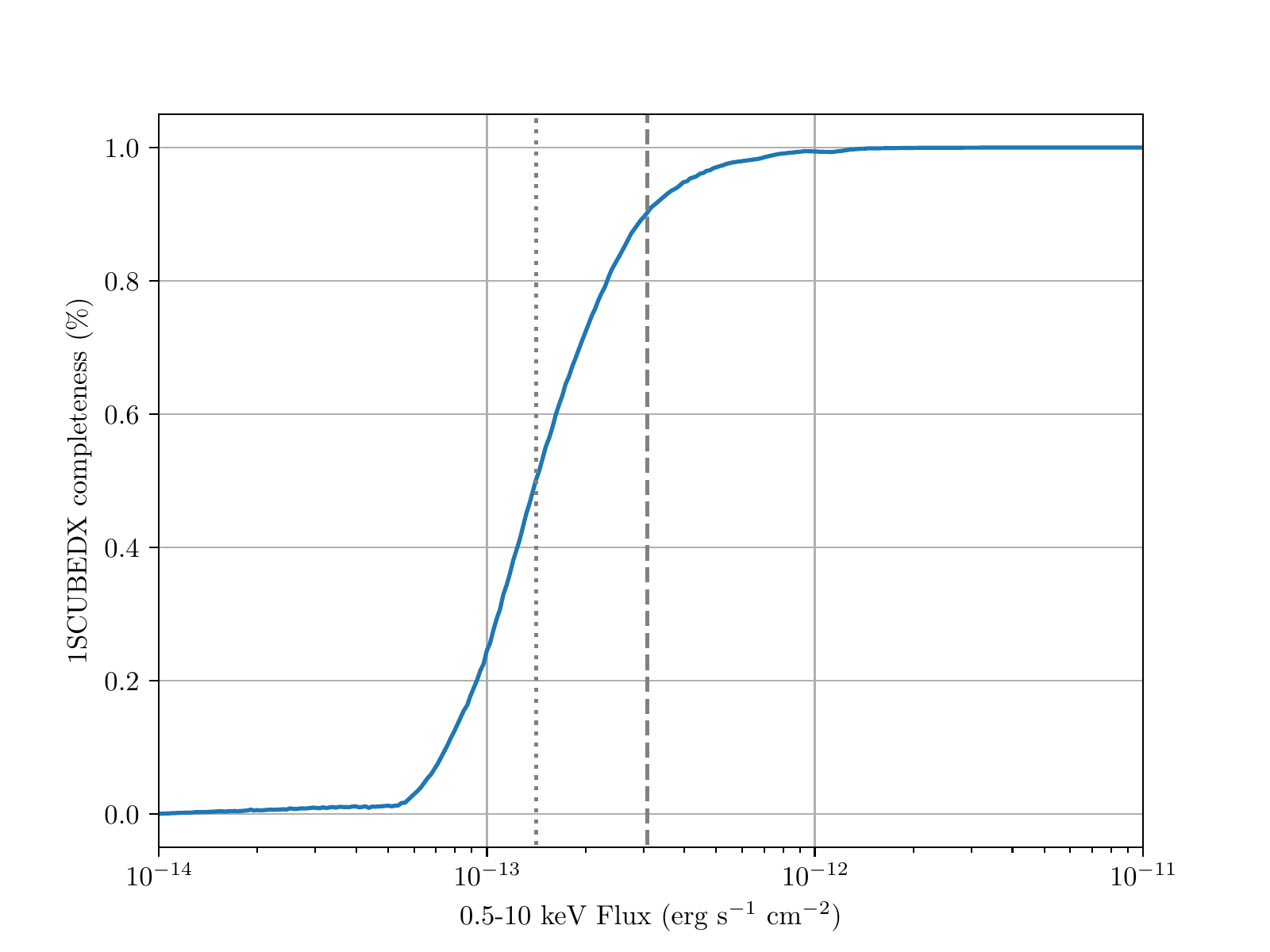}
\caption{Simulated completeness function for the \onex{} survey. The 90\% and 50\% completeness fluxes are marked as dashed and dotted lines respectively.}
\label{fig:completeness}
\end{figure}

In this simulation we estimated the completeness for 500 logarithmically spaced flux levels in the range $10^{-14} - 10^{-11}\ \mathrm{erg\ cm^{-2}\ s^{-1}}$. For each flux level we \editpre{simulated} an X-ray source at random coordinates inside the survey, and \replacedref{calculating}{calculated} the likelihood of its detection. By repeating this over 20,000 trials, we were able to estimate the completeness rate of the \onex{} survey at each flux level.  This completeness rate is shown in Figure~\ref{fig:completeness}. 

These simulations show that the \onex{} catalog is $>50\%$ complete above a flux level of $1.4\times10^{-13}\ \mathrm{erg\ cm^{-2}\ s^{-1}}$ and $>90\%$ complete above a flux level of $3.1\times10^{-13}\ \mathrm{erg\ cm^{-2}\ s^{-1}}$.

\subsection{Luminosity functions}

\begin{figure}[t]
\centering
\includegraphics[width=0.5\textwidth]{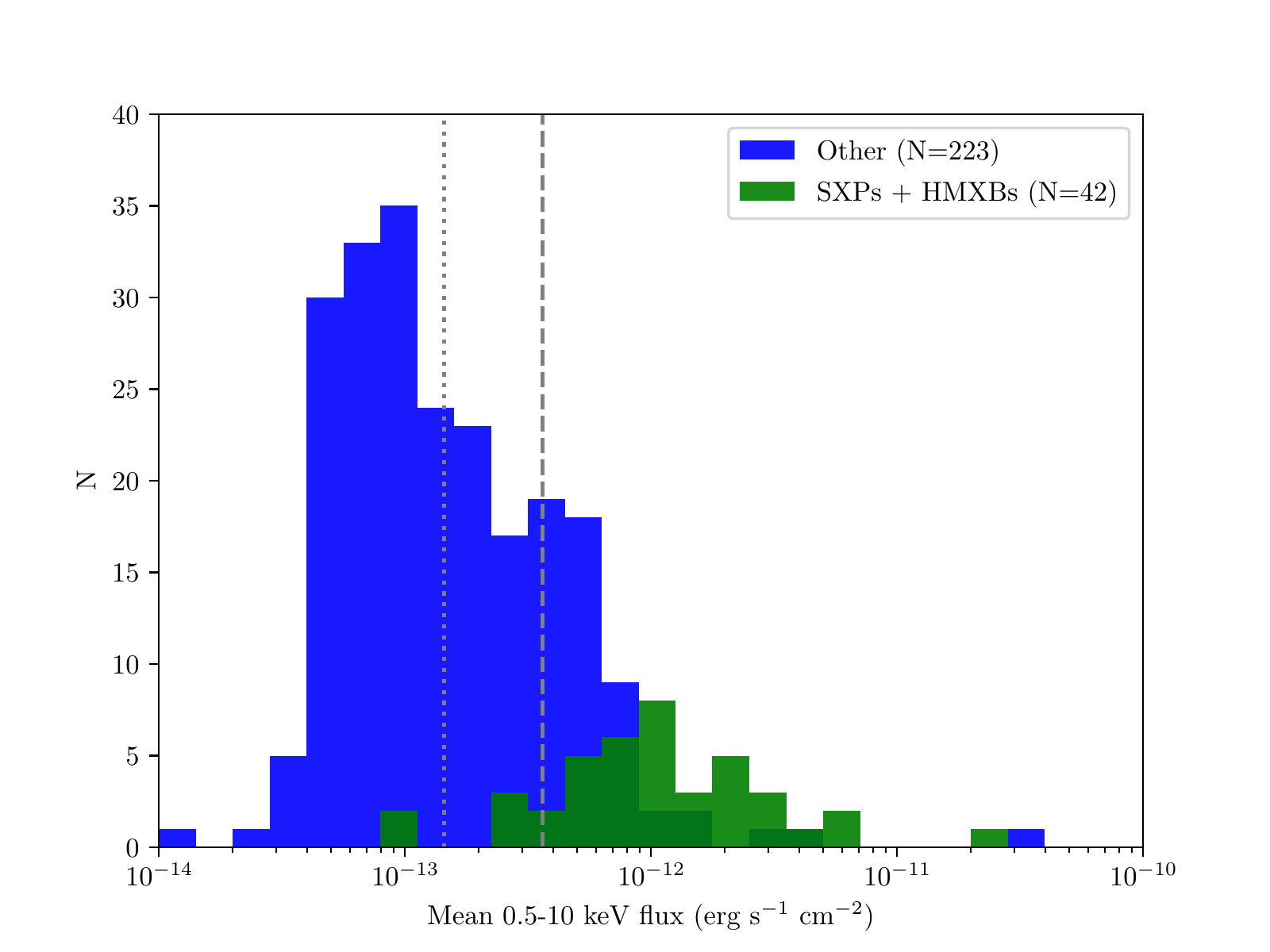}
\caption{Plot showing the histogram of average fluxes rates for all \scubed{} sources flagged as . Green shows sources that are identified HMXBs (including known SXP \bexrbs{} from \citealt{CoeKirk15}), blue shows all other sources in the \onex{} catalog. Fluxes at which the survey are more than 90\% (dashed line) and 50\% (dotted line) complete are shown.}
\label{fig:CRHist25}
\end{figure}

The distributions of the mean fluxes for known HMXBs (including SXPs and other identified HMXBs) and all other \onex{} sources is given in Figure \ref{fig:CRHist25}. It is seen that the average fluxes of the HMXB sources is typically higher than that of the other sources; the median of the average fluxes is $1.15\times10^{-12}\ \mathrm{erg\ s^{-1}\ cm^{-2}}$ for SXP sources and $1.20\times10^{-13}\ \mathrm{erg\ s^{-1}\ cm^{-2}}$ for the unidentified X-ray sources, almost an order of magnitude lower. We note that the majority of the unidentified sources are at mean fluxes below the $90\%$ completeness, suggesting a large population of fainter sources.

To further compare the sources from \scubed{} with other characterizations of the HMXB population within the SMC, a cumulative histogram of the X-ray flux was plotted and a subsequent luminosity function was computed using a least-squares fit to the distribution. 

\begin{figure}[t]
\centering
\includegraphics[width=0.5\textwidth]{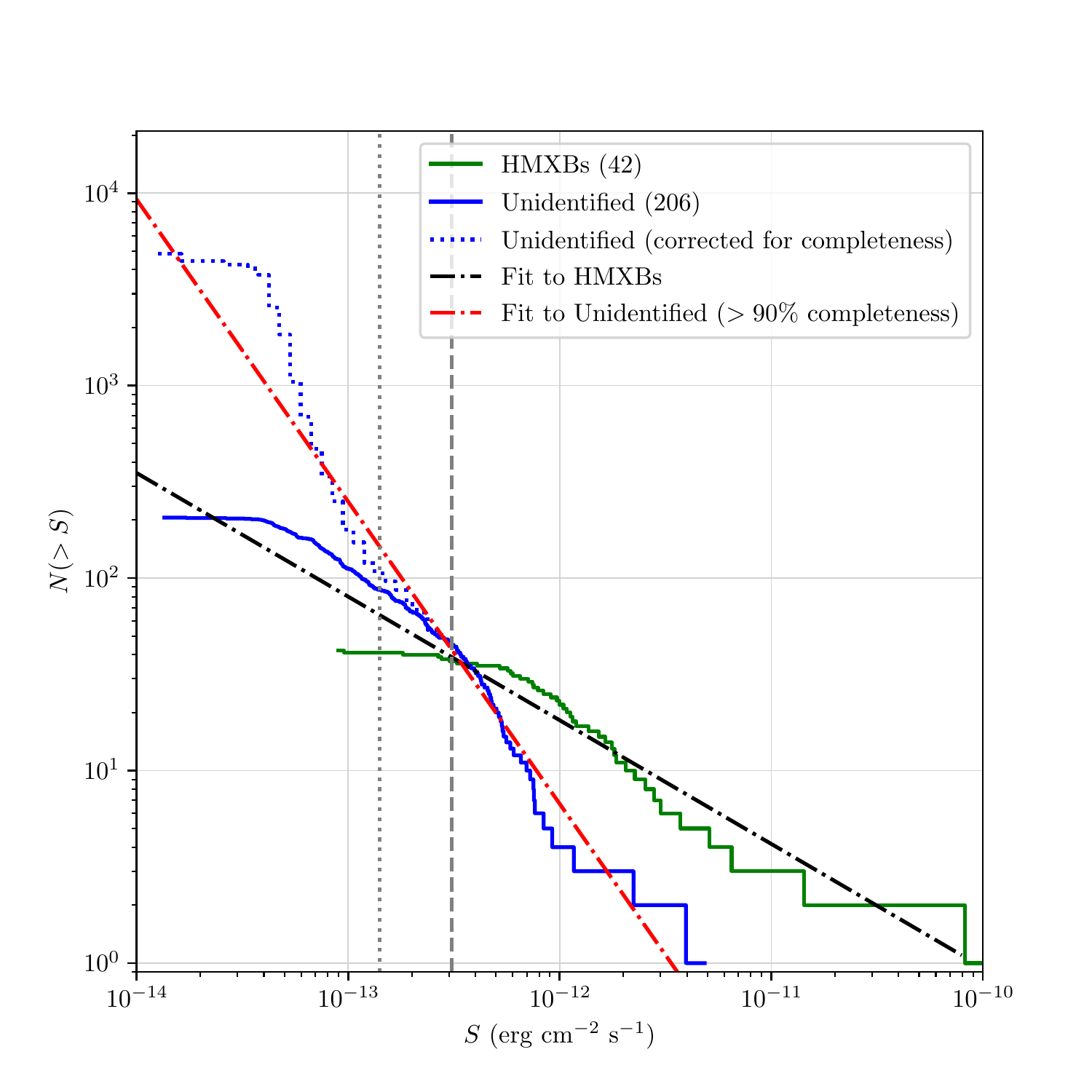}
\caption{logN-logS plot showing the distribution of identified HMXB sources (green) in the \onex{} catalog, compared with the logN-logS distribution of unidentified X-ray sources (blue). Dashed and dotted grey lines show the fluxes at which the \onex{} catalog is expected to be $90\%$ and $50\$$ complete. Straight line fits to the data are indicated, and for unidentified sources, we also plot the logN-logS distribution corrected for completeness.}
\label{fig:LFSXP}
\end{figure}

The results are shown in Figure \ref{fig:LFSXP}.

The least-squares fit to the HMXB distribution in Figure \ref{fig:LFSXP} gives a slope of $\alpha = -0.64 \pm 0.03$. Hence the resulting luminosity function is given by Equation \ref{eqn:LFSXP} where $N_{HMXB}$ is the number of sources with a flux greater than $S$. $S$ is the 0.5--10~keV flux in units of $\mathrm{erg\ s^{-1}\ cm^{-2}}$.

\begin{equation}
N_{\mathrm{HMXB}} = 10^{-6.5 \pm 0.4 } S^{-0.64 \pm 0.03}
\label{eqn:LFSXP}
\end{equation}

 \cite{Shtykovskiy05} quote the universal HMXB luminosity function, given by \cite{Grimm03}, to have a slope of $\alpha_{G} = -0.6$, which is in close agreement with with the \scubed{} derived value. 

A similar histogram was also plotted for the unidentified \scubed{} sources, for which we removed all sources for \deleted{the} which a source type has been previously identified (mostly supernova remnants (SNR), AGN and foreground stars). This can be seen in Figure \ref{fig:LFSXP}. 

The slope for this distribution is seen to be steeper and has a slope of $\alpha = -0.97 \pm 0.06$. However, we note that the cumulative distribution shows significant flattening towards the faint end, likely the signature of the incompleteness of \onex{} at lower flux levels. 

To compensate for this we fitted the cumulative distribution function only for fluxes above $90\%$ completeness, and obtain a steeper $\alpha = 1.57 \pm 0.07$. We also computed a cumulative distribution function for unknown sources corrected for the estimated catalog completion. We note that the fitted luminosity function for fluxes above $90\%$ completion shows good agreement with the estimated distribution below the $90\%$ completeness threshold.

Hence the resulting luminosity function for sources of unknown type is given by Equation~\ref{eqn:LFUnknown}, where $N_{\mathrm{Unidentified}}$ is the number of sources with a luminosity greater than $S$.

\begin{equation}
N_{\mathrm{Unidentified}} = 10^{-18.0 \pm 0.8} S^{-1.57 \pm 0.07} 
\label{eqn:LFUnknown}
\end{equation}

 \cite{Shtykovskiy05} quote results of their luminosity function parameters when all likely HMXB candidates are removed from the data set. According to their results, a slope of $\alpha = 1.48 \pm 0.12$ was obtained when excluding only the likely candidates, and a slope of $\alpha = 1.55 \pm 0.13$ when excluding all possible HMXB candidates. Given the value \replacedref{for}{of} the slope \replacedref{to the}{of} unidentified \scubed{} sources quoted above of $\alpha = 1.57 \pm 0.07$, we find close agreement with between the value for \onex{} unidentified sources, and the quoted values from \cite{Shtykovskiy05}. Our derived value for unidentified sources is also consistent with the expected value for a Euclidean slope ($\alpha = 1.5$), suggesting that the majority of the unidentified source population are likely background sources, such as AGN.

\subsection{Duty Cycles}

The duty cycle (DC) of a HMXB is defined by the fraction of the time that they are in outburst. While physically this may be due to the passage of the NS through the companion's circumstellar disk, an approximation to the DC of a source can be made using:
\begin{equation}
DC = \frac{N_{D}}{N_{T}}
\end{equation}
where $N_{D}$ is the number of times the source was detected over $N_{T}$ total observations, \editpre{based on the assumption that a typical \bexrb{} in the SMC is only detected by \scubed{} when it is undergoing an outburst}. Due to the nature of the exposure times to each tile of the SMC, the DC cannot be taken to be completely representative of the physical nature of X-ray outbursts for SXP sources. 

\begin{figure}[t]
\centering
\includegraphics[width=0.5\textwidth]{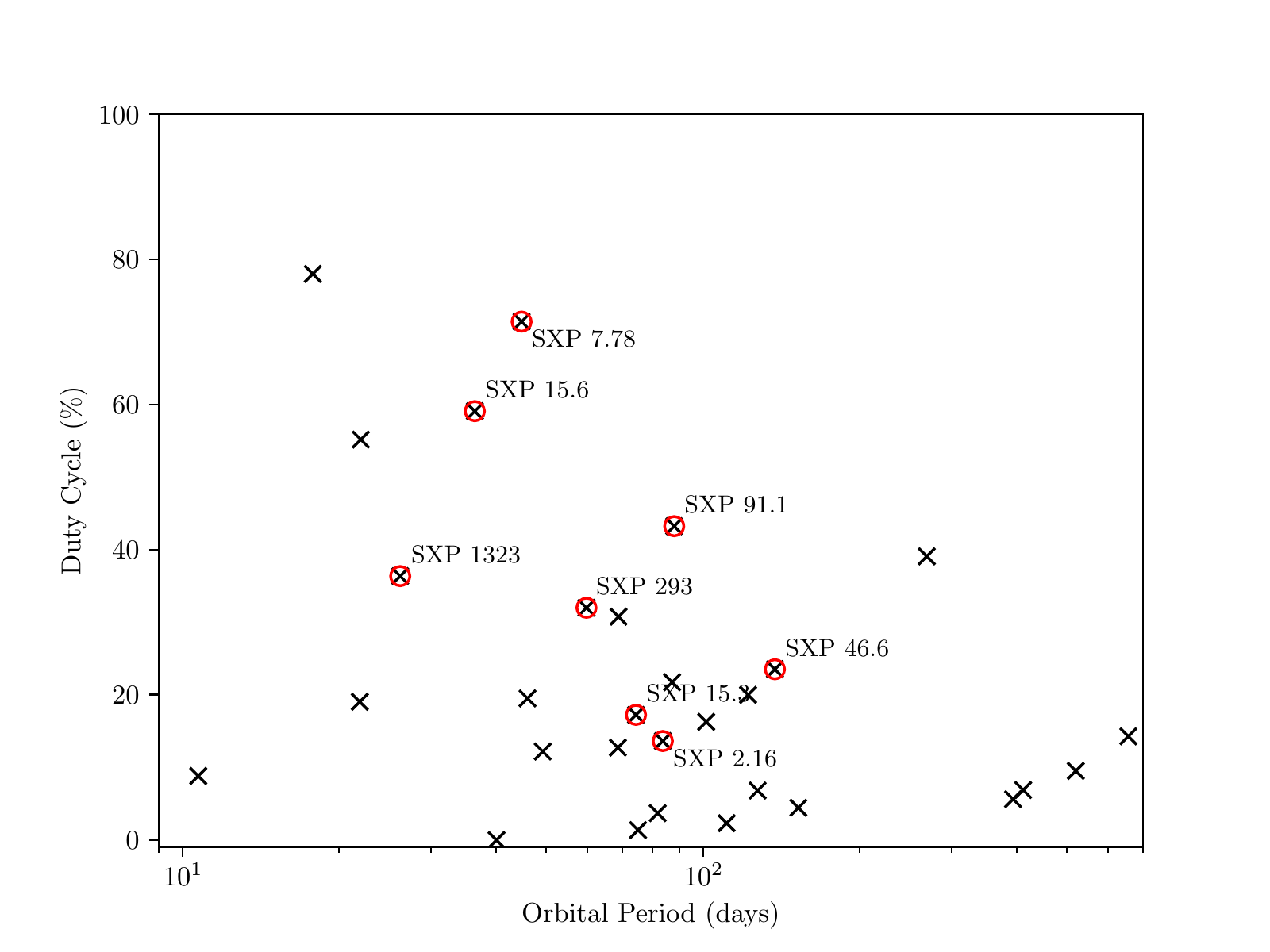}
\caption{Percentage duty cycle of 30 SXP sources against their listed orbital periods. Sources with \scubed{} detected periods from Table~\ref{tab:periods} are highlighted with red circles and labeled with their SXP names. Note that we also highlight SXP~7.78 (= SMC X-3), as the orbital period is detected in this source after the large \typetwo{} outburst as discussed in Section~\ref{sec:smcx3}. 
}
\label{fig:DCvsOrb}
\end{figure}

As the outbursting of the X-ray source is expected to take place during accretion of matter, most probably at periastron, a correlation between the DC of \scubed{} sources and the orbital period of their binary systems was searched for. The catalog of \bexrb{} systems in the SMC \editpre{given in Table 2 of \cite{CoeKirk15}} was used to collect, where available, orbital periods of these binary systems\deleted{ and the NS, respectively}. The plot in Figure \ref{fig:DCvsOrb} shows the DC against \deleted{the} orbital \replacedref{periodicities of the binary system}{period for these 29 systems}. Though they contain a lot of scatter, \deleted{surprisingly,} there is \deleted{clearly} no \replacedref{obvious}{strong} correlation between the X-ray DC and orbital period. \editpre{However, given the approximate nature of utilizing detection fraction as a proxy for DC, it is perhaps not surprising that a strong correlation is not seen, especially given the relatively sparse 7~day observation cadence of \scubed{}, combined with larger observation gaps. This incomplete coverage of SXP sources means that some outbursts may simply have been missed in observation gaps, which could well lead to underestimating the values of DC in some cases using this method.}

To compare the DC with the average flux of each source, the count rates from the light curves were used. For sources with a DC $>$ 0, the average count rate was simply the mean of that for all observations, with non-detections contributing a 0 counts s$^{-1}$ value.

\begin{figure}[t]
\centering
\includegraphics[width=0.5\textwidth]{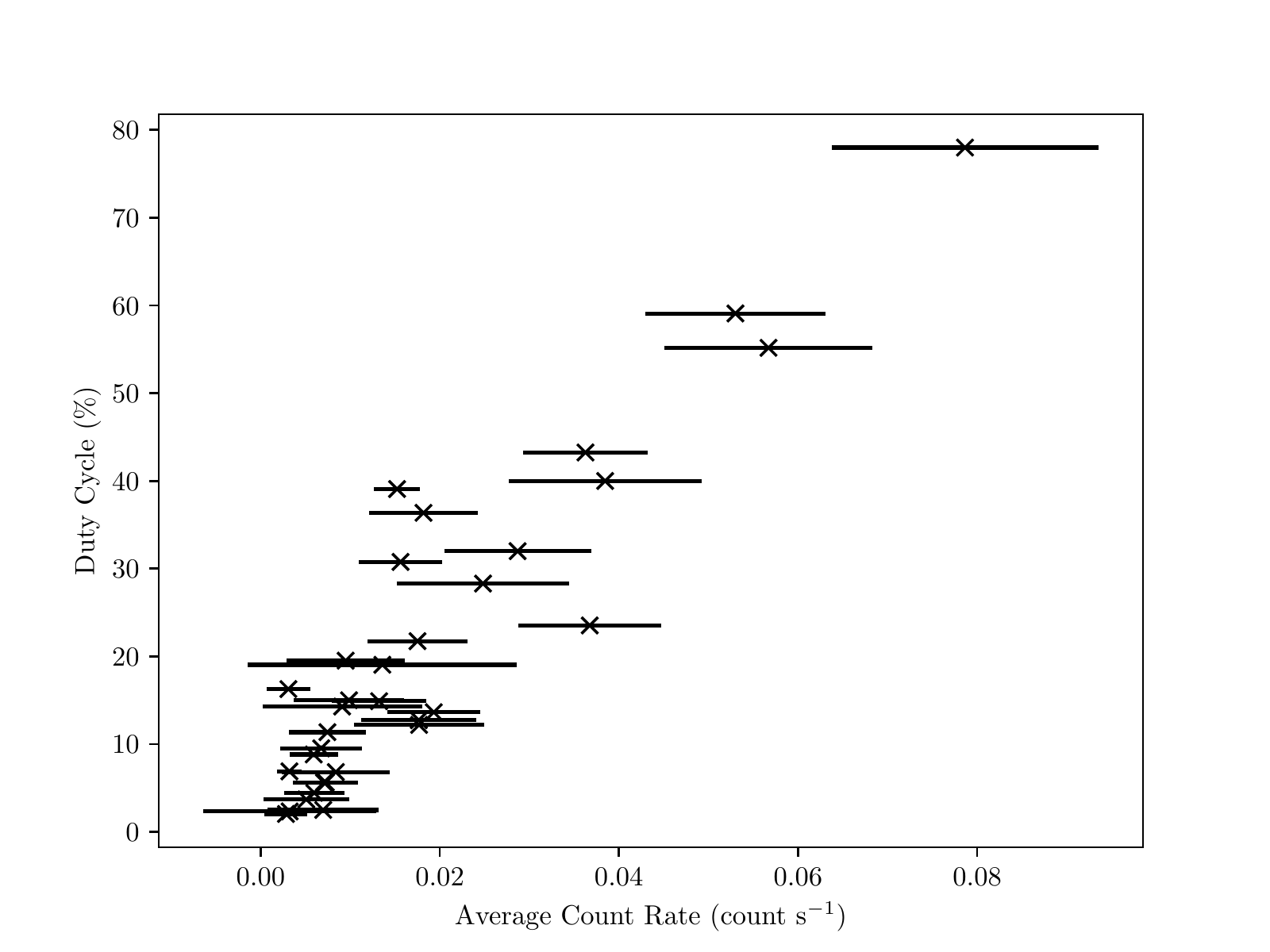} 
\caption{DC of 33 SXP sources, excluding \smcxthree{} and SXP~59.0 which underwent major \typetwo{} outbursts, plotted against their mean count rate across the observation period.}
\label{fig:DCvsCR}
\end{figure}

Figure \ref{fig:DCvsCR} shows the relationship between the DC and the average count rate for each source, as calculated above. If a source was detected during an observation, the mean of the uncertainties in count rate was taken to contribute to the standard error in count rate for each source. The 3$\sigma$ limit for non-detections was used to determine a $1\sigma$ uncertainty to the standard error.

Perhaps not surprisingly, there appears to be a strong correlation between the two variables: a linear regression analysis gave a Pearson correlation coefficient of 
$r=0.918$ and, when considering the size of the data set (32 SXP sources) a p-value of $<<0.001\%$ was obtained, indicating a very strong positive correlation. Therefore the frequency with which a source is detected is much more strongly driven by its average luminosity than any orbital behavior. Other physical parameters were explored, such as the temperature of the companion mass donor star or the proposed NS magnetic field \citep{Klus14}, but none showed a strong a correlation with the DC.

We have shown that we are complete at the 90\% level down to $S_{\mathrm{}}=3\times10^{-13}\ \mathrm{erg\ cm^{-2}\ s^{-1}}$. This corresponds to approx. 0.005 counts/s in XRT. If we look at Figure \ref{fig:DCvsCR} we can see that this 90\% threshold of 0.005 counts/s is on the extreme left of the plot. So all the illustrated DC points are from the brighter sources and hence must all be, at least, 90\% complete/accurate. So we conclude that Figures \ref{fig:DCvsOrb} and \ref{fig:DCvsCR} are not significantly affected by completeness concerns.

\subsection{Confirmation of proposed HMXB systems in \onex{}}

\begin{figure}[t]
\centering
\includegraphics[width=0.5\textwidth]{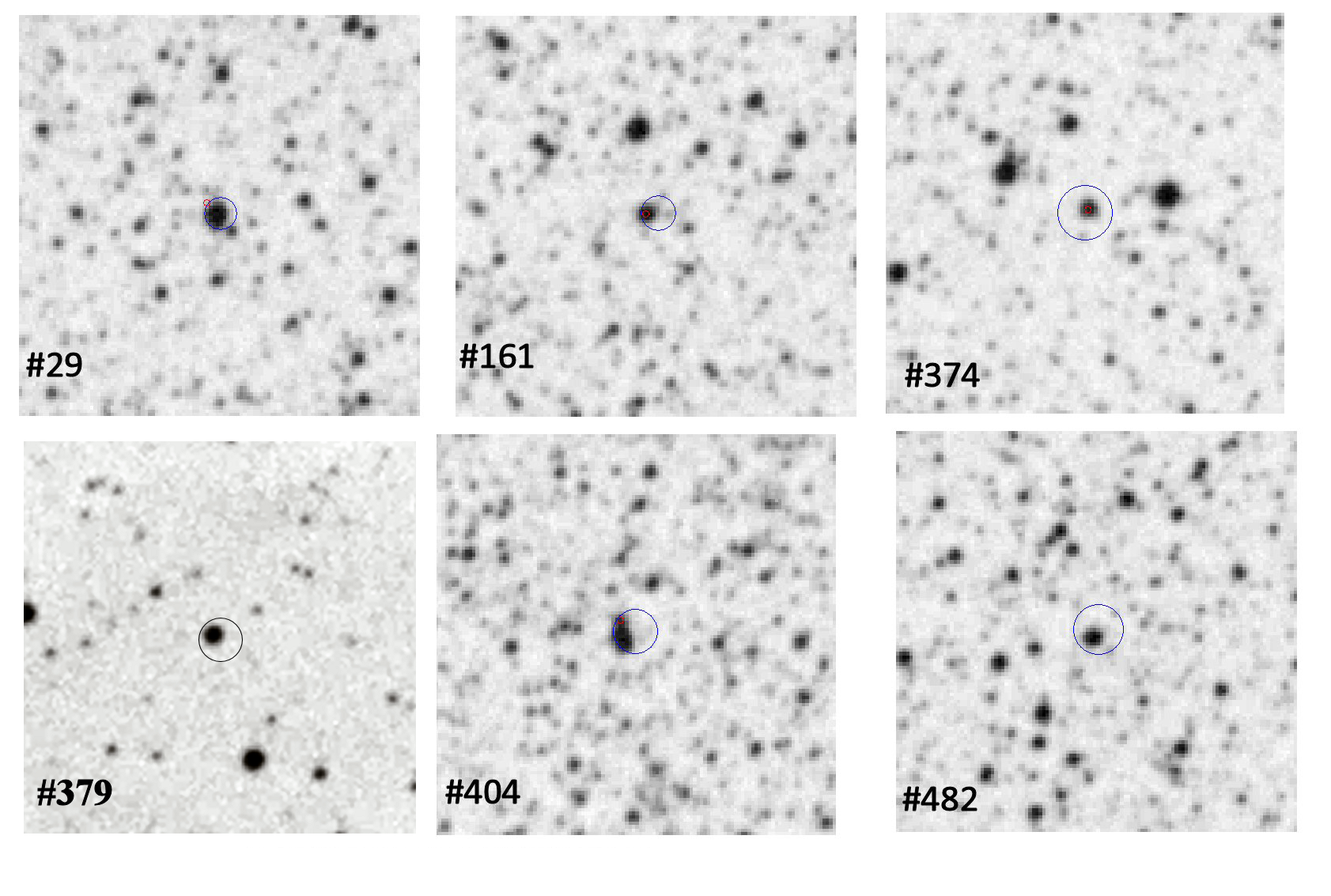}
\caption{\label{fig:SetOf6}Finding charts for the 6 proposed HMXB identifications discussed in the text. Each negative image is taken from the DSS2 red survey and is $2\arcmin \times 2\arcmin$  in size. North is at the top, East to the right. The uncertainty in the position of the \scubed{} source is shown as a blue circle. The red dot indicates the position of the proposed optical counterpart.}
\end{figure}

In order to identify new, or confirm proposed HMXB sources, the list of definite \scubed{} detections was first cross-correlated with the list of proposed and identified HMXBs in \cite{Haberl16}. Their list was produced from their X-ray mapping of the SMC with \xmm{} and contains 148 objects they suggest are, or could be identified HMXBs. To identify a possible match between their catalog and ours it was required that the $1\sigma$ positional uncertainties from the two samples overlapped. The result was 29 matches between the two samples, of which 24 were confirmed SXP sources already listed in \cite{CoeKirk15}. The fields of the remaining 6 objects are shown in Figure \ref{fig:SetOf6} and the details of each of these sources is listed in Table~\ref{tab:new}. From the Figure \ref{fig:SetOf6} it is apparent that there is an optical counterpart present within the \scubed{} positional uncertainty circle in each case. Of the 6 objects listed, the only one that \cite{Haberl16} doubted being identified as an HMXB is SC482. Their doubts arise from the reported lack of H$\alpha$ emission in the proposed optical counterpart. This object is also in the \cite{Evans04} catalog listed as a B0.5V star in the SMC similar to many HMXB systems. In addition, the OGLE III \& IV light-curves \citep{Udalski97,Udalski15} reveal substantial random outbursts from this system with I-band variations as large as 1 magnitude. Such behavior is symptomatic of the growth and decline of circumstellar disk structures around the star. So it is very probable that H$\alpha$ measurements made at the right time would reveal a substantial Balmer line excess and confirm the true nature of this object to be a Be-star and hence this system to be a clear HMXB.

\begin{deluxetable*}{lcccccccc}
    \tablecaption{\label{tab:new}List of proposed non-SXP HMXB systems. The second column refers to matches with the catalog of \cite{Haberl16}. The optical identifications come from the following catalogs: [MA93] \citep{Meyssonnier93}, AzV \citep{Azzopardi79}, [M2002] \citep{Massey02}.}
    \tablewidth{0pt}
    \tablehead{
    \colhead{SC\#} & \colhead{HS\#} & \colhead{Name} & \colhead{RA/Dec (J2000)} &
    \colhead{Err} & \colhead{Det.} & \colhead{Luminosity\dag} & \colhead{Catalog Name} \\
    \colhead{} & & \colhead{(\onex{} J...)} & \colhead{} & \colhead{($''$)} & \colhead{(\%)} & \colhead{(0.5 - 10 keV)}
    }
    \startdata
    SC29 & 125 & J010155.5-723236 & $01^{h} 01^{m} 55\fs53\ {-72}\degr 32\arcmin 36\farcs0$ & $4.9$ & 7.3 & $3.4^{+1.1}_{-1.4} \times 10^{35}$ & AzV 285 \\
    SC161 & 121 & J010029.2-722033 & $01^{h} 00^{m} 29\fs17\ {-72}\degr 20\arcmin 33\farcs1$ & $5.3$ & 11.4 & $1.9^{+0.8}_{-0.8} \times 10^{35}$ & [MA93] 1208 \\
    SC374 & 133 & J010435.6-722149 & $01^{h} 04^{m} 35\fs57\ {-72}\degr 21\arcmin 49\farcs3$ & $8.2$ & 10.0 & $4.9^{+3.1}_{-2.2} \times 10^{35}$ & [MA93] 1470 \\
    SC379 & 143 & J012326.7-732122 & $01^{h} 23^{m} 26\fs67\ {-73}\degr 21\arcmin 22\farcs4$ & $6.7$ & 7.0 & $2.3^{+1.7}_{-1.0} \times 10^{35}$ & [M2002] SMC 81035 \\
    SC404 & 76 & J004929.5-733107 & $00^{h} 49^{m} 29\fs54\ {-73}\degr 31\arcmin 07\farcs8$ & $6.7$ & 4.5 & $4.4^{+2.2}_{-1.1} \times 10^{35}$ & [MA93] 302 \\
    SC482 & 128 & J010331.1-730141 & $01^{h} 03^{m} 31\fs10\ {-73}\degr 01\arcmin 41\farcs2$ & $7.5$ & 7.9 & $1.4^{+0.6}_{-0.9} \times 10^{35}$ & [M2002] SMC 56587 \\
    \enddata
    \tablenotetext{\dag}{Luminosity is a mean value and assumes source is at distance of 62~kpc}
    \end{deluxetable*}
    
More generally, the error circles of all the confirmed \scubed{} objects were searched for matches within the [MA93] \citep{Meyssonnier93} and AzV \citep{Azzopardi79} catalogs. This revealed a total of 29 matches, all of which are known SXP objects. Though this broad search did not reveal any further HMXB candidates, the possibility still remains that some of the \scubed{} objects could still correlate with Be-stars that are not in either, or any, catalog.

In addition to positional matching, we examined the X-ray spectral signatures of these HMXB systems. We found that the mean fitted spectral index for known \bexrbs{} in \onex{} was $\Gamma = 0.9 \pm 0.4$. For the 6 sources listed in Table~\ref{tab:new}, the mean spectral index was $\Gamma = 0.9 \pm 0.5$, i.e. consistent with the population of known \bexrbs{}. In comparison, the average spectral index for all other sources in \onex{} was $\Gamma = 1.4 \pm 0.7$, i.e. softer, although the spread is large enough that we cannot definitely say that these represent a separate population of sources, and cannot rule out the presence of a possible population of unidentified HMXBs. 

\subsection{Orbital Period searches}\label{sec:period_searches}

For each \scubed{} source flagged as `Good', we performed a Lomb-Scargle (L-S) period search (e.g.~\citealt{Lomb76,Scargle82}) of the light-curve data in the range \replacedref{up to 0.5~years}{of 21--180~days}, in order to search for \replacedref{any signs of any}{evidence of} periodicities, perhaps associated with orbital or super-orbital periods. To perform this L-S search, we utilized the implementation present in AstroPy 3.0 \citep{Astropy}, and performed the search on count rate data extracted from the \scubed{} pipeline. In the case of upper-limits, we assumed a count rate of zero, in order to include non-detections in the period search. For each period we calculated the false alarm probability ($P_\mathrm{false}$) for the most significant peak in the periodogram. For the blind period search (i.e. were we do not assume we know what the period is), we only report period detections for $P_\mathrm{false} \le 1\%$, in order to avoid spurious detections. 

In the case of a period detection, we estimate the error on the period measurement using a Monte-Carlo technique \editpre{(e.g.~\citealt{Gotthelf99})}. We created a model of the periodic emission from the \scubed{} light-curve folded at the detected period, \replacedref{The folded light-curve was modeled}{fitted by} by a DC level plus two Gaussians. Utilizing this model, we then generated simulated \scubed{} light-curves, using the same observation times and exposures for each data-point, assuming Poisson statistics. We finally performed a period searches all the simulated light-curves. The quoted $1\sigma$ error is the standard deviation of the measured peak period for 10,000 simulated light-curves. 

Table~\ref{tab:periods} lists all \scubed{} sources with detected periodicities, compared to reported orbital or super-orbital periods found in the literature. Using the blind search ($P_\mathrm{false} \le 1\%$), we find 5 sources with significantly detected periods: SXP~91.1, SMC~X-1, SXP~15.6, SXP~6.85, and SXP~46.6. In all cases the detected periods match within errors previously reported periodicities found from X-ray and optical data (citations given in Table~{tab:periods}), \editpre{except for SXP~6.85 (see Section~\ref{sec:sxp685})}.

We note that $P_\mathrm{false}$ is a measure that such a peak of a given height would occur in an L-S periodogram in the case of a null hypothesis, and as such $P_\mathrm{false}$ is not a valid measure of how likely a detection is to be real \citep{VanderPlas17}, especially in the case where an orbital period is already known to be present. For this reason we also examined L-S periodograms for all sources with a previously published period, looking for peaks in the periodogram near previously reported values. We report in Table~\ref{tab:periods} three sources for which the L-S periodograms peak at periods close to their reported orbital periods, but with relatively low $P_\mathrm{false}$ values: SXP~1323, SXP~2.16 and SXP~293. We believe that these period detections are likely real, despite their low $P_\mathrm{false} \le 1\%$ values, given their consistency with published values. On-going observations of these objects with \scubed{} will likely increase the significance of their detection.

Given that these sources are \bexrbs{}, the likely source of the orbital modulation in these sources are periodic \typeone{} outbursts, that occur during the periastron passages. \replacedref{Searches were performed using a L-S periogram search both as a blind search and targeted search of known periods, which}{An L-S search} was able to detect the signature of these \typeone{} outbursts, even though \scubed{} data are not of sufficient sensitivity or time resolution to resolve the shape of the individual outbursts themselves.

We note that sources with detected periods consistent with previously reported values, lie in the range of 26.2 to 137.4~days. \editpre{Therefore,} non-detections from 9 SXP sources can be explained by the orbital period being too long ($\geq$ 0.5 years) or too short (3 weeks or less) to be detected.  As an example, SXP~18.3, with a reported orbital period of 17.79~days \citep{Coe15}, has a high DC (i.e. it is frequently detected), but no obvious periodic variability has been detected. 

Several SXP sources have frequent detections, and previously reported orbital periods in the expected highest sensitivity range of 21-180 days, but \replacedref{no detection of a period}{no period detection} in \scubed{}. \deleted{SXP~59.0, which may show evidence of a single \typeone{} outburst, but for which the \scubed{} light-curve is dominated by a \typeone{} outburst during the period when a 2nd periastron passage would occur, making detection of the relatively long (122~day) orbital period not possible.} \editpre{For example} SXP~327, SXP~967, SXP~169, SXP~175 and SXP~264 all have orbital periods in the right range, but no detected period. 

The lack of a period detection in SXP~175 and SXP~172 \replacedref{are}{is} especially surprising, given that their orbital periods, \editpre{87.2 and 68.8~days respectively}, lie in the middle of the highest sensitivity period range, and have DCs of 22\% and 32\%, which places them between other sources that have period detections on Figure~\ref{fig:DCvsOrb}.

In the case of SXP~172, the orbital period has been previously been measured in X-ray \deleted{by \rxte{}, during a series of periodic outbursts seen} by \rxte{} in data taken frequently between 1999 and 2009 \citep{Schurch11}. In \rxte{} data these outbursts peak between $\sim 0.1 -- 0.5$ PCA counts PCU$^{-1}$ s$^{-1}$, which assuming a type \bexrb{} spectrum, converts to a count rate of $\sim0.04$ XRT count s$^{-1}$, or around 2.3 counts in 60~s exposure. Therefore, it is possible that \typeone{} bursts are too faint to detect in \scubed{} data.

\begin{deluxetable*}{lccccccc}[t]
    \tablecaption{\label{tab:periods}Table of detected periodicities in \scubed{} data, compared with known periodicities in sources. We note that most of these periods are consistent with published values, however the reported period for SXP~6.85, is discrepant from the value found in the literature, and unlikely to be related to the orbital period. Table is ordered from most significant detection to least.}
    \tablewidth{0pt}
    \tablehead{\colhead{SC\#} & \colhead{Name} & \colhead{Catalog} & \colhead{Period} & Orbit & \colhead{$P_\mathrm{false}$} & \colhead{Ref.}\\
     & \colhead{\onex{} J...} & \colhead{Name} & \colhead{(days)} & (days) & (\%) & }
    \startdata
     SC6  & J005056.4-721333 & SXP~91.1     & $89.25 \pm 2.54$ & $88.37\pm0.03$  & $1.9\times10^{-5}$ & [1]  \\
     SC1  & J011705.2-732635 & SMC X-1      & 53.38       & $\sim55$\dag   & $8.0\times10^{-4}$ & [2]  \\
     SC3  & J004854.9-734945 & SXP~15.6     & $36.82 \pm 1.53$ & $36.43 \pm 0.01$ & $2.6\times10^{-2}$ & [3]  \\
     SC49 & J010252.2-724433 & SXP~6.85     & $161.55 \pm 5.28$ & $21.9\pm0.1$   & $3.5\times10^{-2}$ & [4]   \\
     SC271 & J005354.9-722646 & SXP~46.6     & $143.29 \pm 4.5$ & $137.4\pm0.4$  & 1.0        & [5]  \\
     SC11 & J010336.0-720130 & SXP~1323     & $25.80 \pm 0.43$ & $26.188\pm0.045$ & 7.9        & [6]  \\
     SC16 & J012140.6-725731 & SXP~2.16     & $80.25 \pm 0.40$* & $82.5\pm0.05$  & 19         & [7]  \\
     SC10 & J005811.2-723051 & SXP~293      & $59.62 \pm 0.95$ & $59.73\pm0.01$  & 73.8        & [1]  \\
    \enddata
    \tablecomments{References for orbital periods: [1] \cite{Bird12}, [2] \cite{Trowbridge07} - note that SMC~X-1 is quasi-periodic, so the mean period is quoted here, [3] \cite{McBride17}, [4] \cite{Townsend13} , [5] \cite{Galache08}, [6] \cite{Carpano2017}, [7] \cite{Boon17} - reported period is from BAT data.}
    \tablenotetext{\dag}{Note that SMC X-1 is reported to be quasi-periodic in X-rays \citep{Trowbridge07}, in this case we quote the SCUBED peak periodicity without an associated error, as the Monte-Carlo simulation would not be valid for an object in which the period is not constant.}
    \tablenotetext{*}{Error is likely underestimated for SXP~2.16, due to apparent turn-off of X-ray activity from this source after the final detection on 2016 October 5.}
    \end{deluxetable*}

\deleted{The bright and highly variable source SMC~X-1, which shows strong quasi-periodic super-orbital variability was detected by \scubed{} in every observation, and comparison between the \scubed{} data and the BAT transient monitor detections of the source shows that the \scubed{} flux levels and the 15-150 keV BAT detected levels are highly consistent. }

\section{Sources of Interest}\label{sec:results}

\movetabledown=1.50in
 \begin{deluxetable*}{lllccrrrrr}
  \tabletypesize{\scriptsize}
  \tablecolumns{10}
\tablecaption{\label{tab:scubedsources} Sources of interest in the \scubed{} catalog, ordered by date of first detection. 
In this case interesting sources are defined as those that are frequently ($>20\%$ of observations) detected or highly 
variable ($\chi^2_\mathrm{red}>2$ and detected in $>10\%$ of observations).}
\tablehead{
\colhead{SC\#} & 
\colhead{Name} & 
\colhead{First Detection} & 
\colhead{RA/Dec (J2000)} &
\colhead{Err} & 
\colhead{Det.} & 
\colhead{Luminosity\dag} & 
\colhead{Catalog Name} & 
\colhead{Type} &
\colhead{Ref.\ddag}
\\
\colhead{} & 
\colhead{(1SCUBEDX J...)} & 
\colhead{} & 
\colhead{} & 
\colhead{($''$)} &
\colhead{($\%$)} & 
\colhead{(0.5 - 10 keV)} & 
\colhead{} & 
\colhead{} &
\colhead{}
}
\startdata
 SC1  & J011705.2-732635 & 2016 June 08   & $01^{h} 17^{m} 05\fs18\ {-73}^\circ 26' 35\farcs8$ & $2.2^*$ & 100.0 & $1.7^{+0.1}_{-0.1} \times 10^{38}$  & SMC X-1  & HMXB  & [1] \\
 SC2  & J010401.3-720155 & 2016 June 09   & $01^{h} 04^{m} 01\fs29\ {-72}^\circ 01' 55\farcs7$ & $3.8$  & 100.0 & --                  & 1E~0102.2$-$7219 & SNR &  \\
 SC3  & J004854.9-734945 & 2016 June 24   & $00^{h} 48^{m} 54\fs93\ {-73}^\circ 49' 45\farcs6$ & $3.8$  & 59.1 & $1.9^{+0.4}_{-0.3} \times 10^{36}$  & SXP 15.6  & \bexrb{}& [1,2]\\
 SC5  & J004910.3-724939 & 2016 June 24   & $00^{h} 49^{m} 10\fs29\ {-72}^\circ 49' 39\farcs2$ & $4.1$  & 78.0 & $3.8^{+0.6}_{-0.5} \times 10^{36}$  & SXP 18.3  & \bexrb{}& \\
 SC6  & J005056.4-721333 & 2016 June 24   & $00^{h} 50^{m} 56\fs39\ {-72}^\circ 13' 33\farcs4$ & $4.6$  & 43.2 & $1.3^{+0.4}_{-0.3} \times 10^{36}$  & SXP 91.1  & \bexrb{}& \\
 SC8  & J003235.5-730650 & 2016 June 28   & $00^{h} 32^{m} 35\fs46\ {-73}^\circ 06' 50\farcs9$ & $2.3^*$ & 71.4 & $1.5^{+0.1}_{-0.2} \times 10^{-12}$ &      & AGN  & [3]\\
 SC10 & J005811.2-723051 & 2016 June 28   & $00^{h} 58^{m} 11\fs15\ {-72}^\circ 30' 51\farcs0$ & $4.6$  & 32.0 & $1.1^{+0.3}_{-0.3} \times 10^{36}$  & SXP 293  & \bexrb{}& \\
 SC11 & J010336.0-720130 & 2016 July 04   & $01^{h} 03^{m} 36\fs00\ {-72}^\circ 01' 30\farcs8$ & $5.4$  & 36.4 & $7.8^{+2.8}_{-2.3} \times 10^{35}$  & SXP 1323  & \bexrb{}& \\
 SC13 & J005455.4-724513 & 2016 July 04   & $00^{h} 54^{m} 55\fs42\ {-72}^\circ 45' 13\farcs1$ & $4.0$  & 39.1 & $6.3^{+1.2}_{-1.0} \times 10^{35}$  & SXP 504  & \bexrb{}& \\
 SC16 & J012140.6-725731 & 2016 July 06   & $01^{h} 21^{m} 40\fs63\ {-72}^\circ 57' 31\farcs9$ & $3.2^*$ & 13.6 & $7.4^{+3.7}_{-1.8} \times 10^{35}$  & SXP 2.16  & \bexrb{}& [4]\\
 SC17 & J010428.3-723135 & 2016 July 06   & $01^{h} 04^{m} 28\fs32\ {-72}^\circ 31' 35\farcs3$ & $5.1$  & 28.3 & $1.0^{+0.2}_{-0.2} \times 10^{36}$  & SXP 707  & \bexrb{}& \\
 SC20 & J005919.8-722317 & 2016 July 06   & $00^{h} 59^{m} 19\fs82\ {-72}^\circ 23' 17\farcs9$ & $4.3$  & 40.0 & $1.4^{+0.3}_{-0.2} \times 10^{36}$  & SXP 202A  & \bexrb{}& [5] \\
 SC32 & J011838.2-732533 & 2016 August 02  & $01^{h} 18^{m} 38\fs21\ {-73}^\circ 25' 33\farcs7$ & $7.1^*$ & 47.5 & $6.0^{+20.8}_{-2.3} \times 10^{-12}$ & HD 8191A/B  & FG-star& \\
 SC49 & J010252.2-724433 & 2016 August 02  & $01^{h} 02^{m} 52\fs17\ {-72}^\circ 44' 33\farcs4$ & $4.4$  & 55.2 & $2.3^{+0.4}_{-0.3} \times 10^{36}$  & SXP 6.85  & \bexrb{}& [1]\\
 SC71 & J005205.3-722603 & 2016 August 02  & $00^{h} 52^{m} 05\fs27\ {-72}^\circ 26' 03\farcs8$ & $4.3$  & 71.4 & $1.9^{+0.2}_{-0.1} \times 10^{37}$  & SMC X-3  & \bexrb{}& [6,7,8]\\
 SC80 & J005052.5-710902 & 2016 August 02  & $00^{h} 50^{m} 52\fs47\ {-71}^\circ 09' 02\farcs3$ & $6.8$  & 59.5 & --                  & HD 5028  & FG-star&\\
 SC143 & J003108.2-731207 & 2016 August 04  & $00^{h} 31^{m} 08\fs18\ {-73}^\circ 12' 07\farcs3$ & $4.1$  & 20.8 & $1.1^{+2.0}_{-0.6} \times 10^{36}$  &      & Unknown \\
 SC148 & J005151.6-731031 & 2016 August 08  & $00^{h} 51^{m} 51\fs56\ {-73}^\circ 10' 31\farcs5$ & $7.4$  & 30.8 & $4.9^{+4.5}_{-1.1} \times 10^{35}$  & SXP 172  & \bexrb{}&\\
 SC160 & J010151.2-722334 & 2016 August 09  & $01^{h} 01^{m} 51\fs19\ {-72}^\circ 23' 34\farcs6$ & $5.0$  & 21.7 & $6.0^{+1.5}_{-1.3} \times 10^{35}$  & SXP 175  & \bexrb{}&\\
 SC271 & J005354.9-722646 & 2016 September 22 & $00^{h} 53^{m} 54\fs94\ {-72}^\circ 26' 46\farcs1$ & $4.4$  & 23.5 & $1.2^{+0.2}_{-0.2} \times 10^{36}$  & SXP 46.6  & \bexrb{}&\\
 SC372 & J005212.9-731916 & 2016 November 24 & $00^{h} 52^{m} 12\fs88\ {-73}^\circ 19' 16\farcs7$ & $4.8$  & 17.2 & $8.8^{+0.5}_{-0.6} \times 10^{36}$  & SXP 15.3  & \bexrb{}&\\
 SC403 & J005456.4-722647 & 2016 December 09 & $00^{h} 54^{m} 56\fs43\ {-72}^\circ 26' 47\farcs7$ & $4.0$  & 20.0 & $3.1^{+0.4}_{-0.4} \times 10^{36}$  & SXP 59.0  & \bexrb{}&[9]\\
 \enddata
\tablenotetext{\dag}{Luminosity assumes source is at distance of 62~kpc and are given in the energy range of 0.5--10~keV. For sources SC8 (AGN), and HD 8191A/B (foreground star), distances are unknown so fluxes (units: $\mathrm{erg\ s^{-1}\ cm^2}$) are given. SC80 AKA HD 5028 is contaminated by optical loading, so no luminosity is given. 1E~0102.2-7219 (SC2) is an extended source, so the \scubed{} calculated luminosity is not correct, and therefore is omitted.}
\tablenotetext{\ddag}{References:
 [1] \cite{KenneaATel9299}, 
 [2] \cite{EvansATEL9197}, 
 [3] \cite{CoeATEL9414}, 
 [4] \cite{Boon17}, 
 [5] \citep{CoeATEL9307}, 
 [6] \cite{KenneaATel9362}, 
 [7] \cite{KenneaATel9677}, 
 [8] \cite{Townsend2017}, 
 [9] \cite{KenneaATel10250}. 
} 
\tablenotetext{*}{Position is enhanced utilizing UVOT to correct for systematic errors in astrometery using the method of \cite{Goad07}.}
\end{deluxetable*}

In this section we report on sources of particular interest from the first year of monitoring. We define interesting sources as those that are detected in $>20\%$ of all observations, and those that are considered highly variable. \replacedref{The measure of variability}{Variability} in this case is Pearson's $\chi^2$, i.e. we fit a model of a constant level to each \scubed{} light-curve (points are weighted by their measurement errors). We then find a reduced $\chi^2$ ($\chi^2_\mathrm{red}$) for the constant level fit, and $\chi^2_\mathrm{red} > 2$ is considered to be highly variable. These sources of interest are given in Table~\ref{tab:scubedsources}. In this table, we also present a mean luminosity level for each source, as estimated by performing a simple power-law fit to the spectrum of the combined \scubed{} observations, correcting for absorption, and assuming a standard SMC distance of 62~kpc. The majority of the ``Sources of Interest'' based on these criteria are previously identified \bexrb{}.

\subsection{Uncataloged Sources}

Two sources have been found to be frequently ($>20\%$ of the time) detected by \scubed{} which are previously uncataloged X-ray emitters: \onex{}~J003235.5-730650 (SC8) and \onex{}~J003108.2-731207 (SC143). In this section we take a closer look at these two objects, in order to determine their source type.

\subsubsection{\onex{}~J003235.5$-$730650}\label{sec:agn}

\scubed{} detected \onex{}~J003235.5$-$730650 (SC8) in $71.8\%$ of all observations performed in the first year of observations. The \swift/XRT derived position does not match any known cataloged X-ray point source, \replacedref{in addition the Swift localization does not match}{or} any known optical sources. In order to investigate the nature of this object, TOO observations were performed by \swift{} for a total exposure of 12.3~ks between 2016 July 7 and 2016 August 29, and also \nustar{} \editpre{(Observation ID 90201030002)} on 2016 July 17, for $\sim55$~ks, and \chandra{} \editpre{(Observation ID 19691)} on 11 August 2016 for $\sim1$~ks. Analysis of these observations has been previously reported by \cite{CoeATEL9414}. 

The \nustar{} observation was performed primarily to investigate the possibility that this source was either an accreting pulsar or magnetar, by searching for any pulsar periodicity. However, analysis of the \nustar{} data does not reveal the presence of any significant temporal variations. Spectral analysis of the \nustar{} data reveal that the source is well fit using an power-law model (XSPEC \texttt{tbabs * power} model), with a photon index of $1.65^{+0.15}_{-0.10}$ with a \nustar{} flux of $2 \times 10^{-12}\ \mathrm{erg\ s^{-1} cm^{-2}}$ (3 - 78 keV), given \nustar{}'s 3 keV low energy cut-off, the absorption is not well constrained for these data. The averaged spectrum from combined \scubed{} data, was well fit by a photon index of $1.53^{+0.21}_{-0.09}$, consistent within errors with the \nustar{} spectrum, the absorption is consistent with the Galactic value of $1.93 \times 10^{21}\ \mathrm{cm^{-2}}$ \citep{Willingale13}. 

The lack of optical counterpart consistent with the XRT derived position was an enigma. Analysis of the TOO XRT data revealed a large disparity ($13\farcs5$) between the XRT position derived using XRT only data, and the position utilizing UVOT data to correct for astrometric errors \citep{Goad07}. This disparity motivated a short $\sim1$~ks observation with \chandra{}, in order to obtain a better localization of the source. The \chandra{} position for \onex{}~J003235.5$-$730650 was found to be RA/Dec(J2000) = $00^h 32^m 34\fs72 -73\degr 06\arcmin 49\farcs14$. This position lies $3\farcs7$ from the best \scubed{} derived position given in Table~\ref{tab:scubedsources}, outside of the $2\farcs3$ radius $90\%$ confidence error radius. We note that the \chandra{} position is consistent with the XRT-only derived position from the TOO data. This new position is consistent with the location of an optical source seen in OGLE III and OGLE IV data, with an apparent non-stellar PSF \citep{CoeATEL9414}, suggesting that this source is likely a background AGN. We also note that this object is in the AllWISE AGN catalog \citep{Secrest15}, and that the power-law spectrum detected by \swift{} and \nustar{} is consistent with the AGN hypothesis.

\clearpage
\subsubsection{\onex{}~J003108.2$-$731207}

\onex{}~J003108.2$-$731207 (SC143) was detected in $20.8\%$ of \scubed{} observations and has a mean flux of $2.3^{+4.3}_{-1.5} \times 10^{-13}\ \mathrm{erg s^{-1} cm^{-2}}$ (0.5 - 10 keV, correct for absorption). The average spectrum combining all \scubed{} data reveals a soft spectrum, with a power-law fit giving a photon index of $4.8^{+1.8}_{-1.2}$, suggesting that the source is not a \bexrb{}, which typically have hard ($\Gamma \simeq 1.0$) spectra. However, the power-law fit also requires a high absorption ($0.49^{+0.33}_{0.22} \times 10^{22}\ \mathrm{cm^{-2}}$, which given the low line-of-sight absorption in the SMC would require the absorption to be localized to the source. As an alternative, fitting a thermal (XSPEC's \texttt{apec}) model provides a good fit to the data with $N_\mathrm{H}$ fixed at the expected SMC value, with $\chi^2_\mathrm{red} = 0.924$ (13 dof), which is an improved fit over the absorbed power-law model ($\chi^2_\mathrm{red} = 1.126$ for 13 dof). This gives a fitted $kT = 0.74^{+0.19}_{-0.18}$~keV, with an metallic abundance of $0.13^{+0.23}_{-0.09}$.

The fitted average flux of \onex{}~J003108.2$-$731207 is $1.06^{+0.41}_{-1.01} \times 10^{-13}\ \mathrm{erg\ s^{-1}\ cm^{-2}}$ (0.5 - 10 keV), equivalent to $\sim4.5 \times 10^{34}\ \mathrm{erg s^{-1}}$ at 62~kpc.

A catalog search reveals no known X-ray point sources at or near the \scubed{} position. However, we note that this source is inside the same pointing as source SC8, for which the \scubed{} derived position was not consistent with a \chandra{} position. Unfortunately, SC143 is outside of the field of view of the \chandra{} observation, however an XRT-only position derived from TOO observations was found to be consistent with the \chandra{} localization, so we utilized these data to calculate an updated position for SC143, which was found to be RA/Dec(J2000) = $00^h 31^m 08.59^s -73\degr 12\arcmin 06\farcs4$ with an estimated error of 4 arcseconds (90\% confidence). 

A catalog search reveals several possible optical and IR counterparts at this position. A nearby bright (J=11.651) source 2MASS~00310958-7312082 \citep{Cutri03} lies 4.7 arcseconds from the best XRT position, just outside the error circle, is likely a good candidate to be the optical counterpart of SC143. However, \cite{Kato07} lists 2 fainter (J=19.08 and J=18.88) point sources inside the XRT error circle that could also be associated with the source. Given the uncertainty with the astrometry in this field, it is clear that we cannot definitively associate an optical/IR counterpart with this source. Therefore, further study of this source with deeper X-ray observations, and a positive identification of the optical/IR counterpart will be required in order to classify the source type for SC143.

\subsection{Sources showing significant outbursts or variability}

\begin{figure*}
\centering
\includegraphics[width=1.0\textwidth]{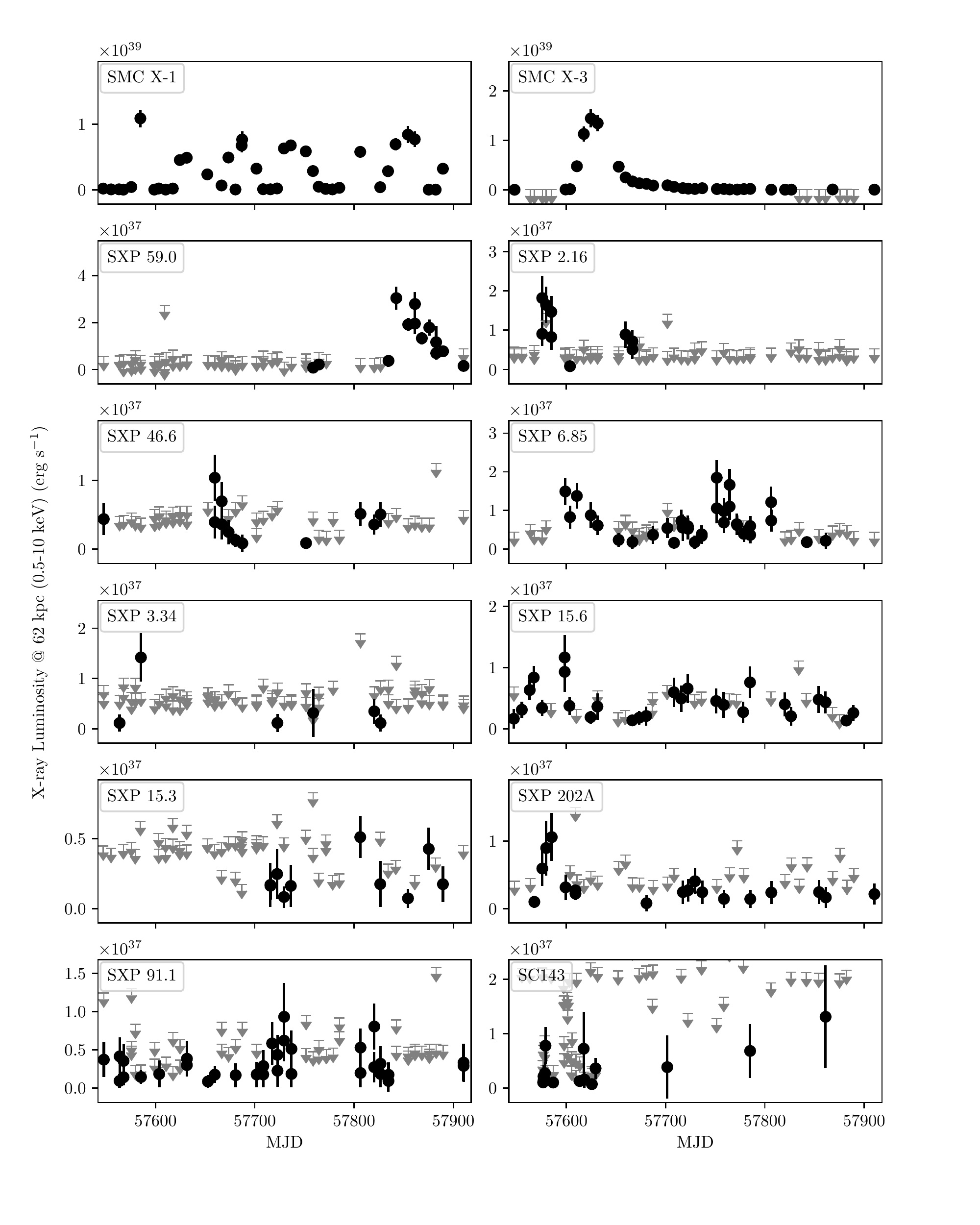}
\caption{\label{fig:flaring}\scubed{} light-curves for flaring and variable sources in the SMC during the first year of \scubed{}. Upper-limits for non-detections are shown in light gray.}
\end{figure*}

\deleted{In order to determine if a source was in outburst, we fit a mean level to each \scubed{} light-curve, and calculated a reduced $\chi^2$ value in order to statistically determine the magnitude of deviation from a simple constant. For this fit, we used only datapoints where \scubed{} detected the point source, and only report on sources that have greater than 4 detections during the period of interest.} Figure~\ref{fig:flaring} shows light-curves of the sources in Table~\ref{tab:scubedsources} that show statistically significant (reduced $\chi^2 > 2.0$ when fit with a constant model) degree of variability. \deleted{(for $\chi^2$ calculation, the measured $1\sigma$ errors on count rates are used to as weights)}. \deleted{Luminosity values assume a fixed SMC distance of 62~kpc, and utilizes a counts-to-flux conversion derived from the mean \scubed{} spectrum for each source based on the combined total exposure over the first year of observations.} In the following subsections, we discuss the details of individual outbursting sources, including analysis of TOO observations from Swift and other observatories, if those results have not been previously published.

\subsubsection{SMC~X-1\label{sec:smcx1}}

SMC X-1 is a HMXB containing an X-ray pulsar with a 0.71~s period, the companion star is a B0 supergiant (e.g.~\citealt{Li97}). SMC X-1 is a persistent X-ray emitter, but shows large quasi-periodic super-orbital variations with an average period of $\sim55$~days \citep{Trowbridge07}, which can be clearly seen in Figure~\ref{fig:flaring}. SMC X-1 is the only variable source in \scubed{} survey to be detected in all observations.
The \scubed{} detection of SMC~X-1 was previously reported by \cite{KenneaATel9299}. 

\begin{figure}[t]
\centering
\includegraphics[width=0.5\textwidth]{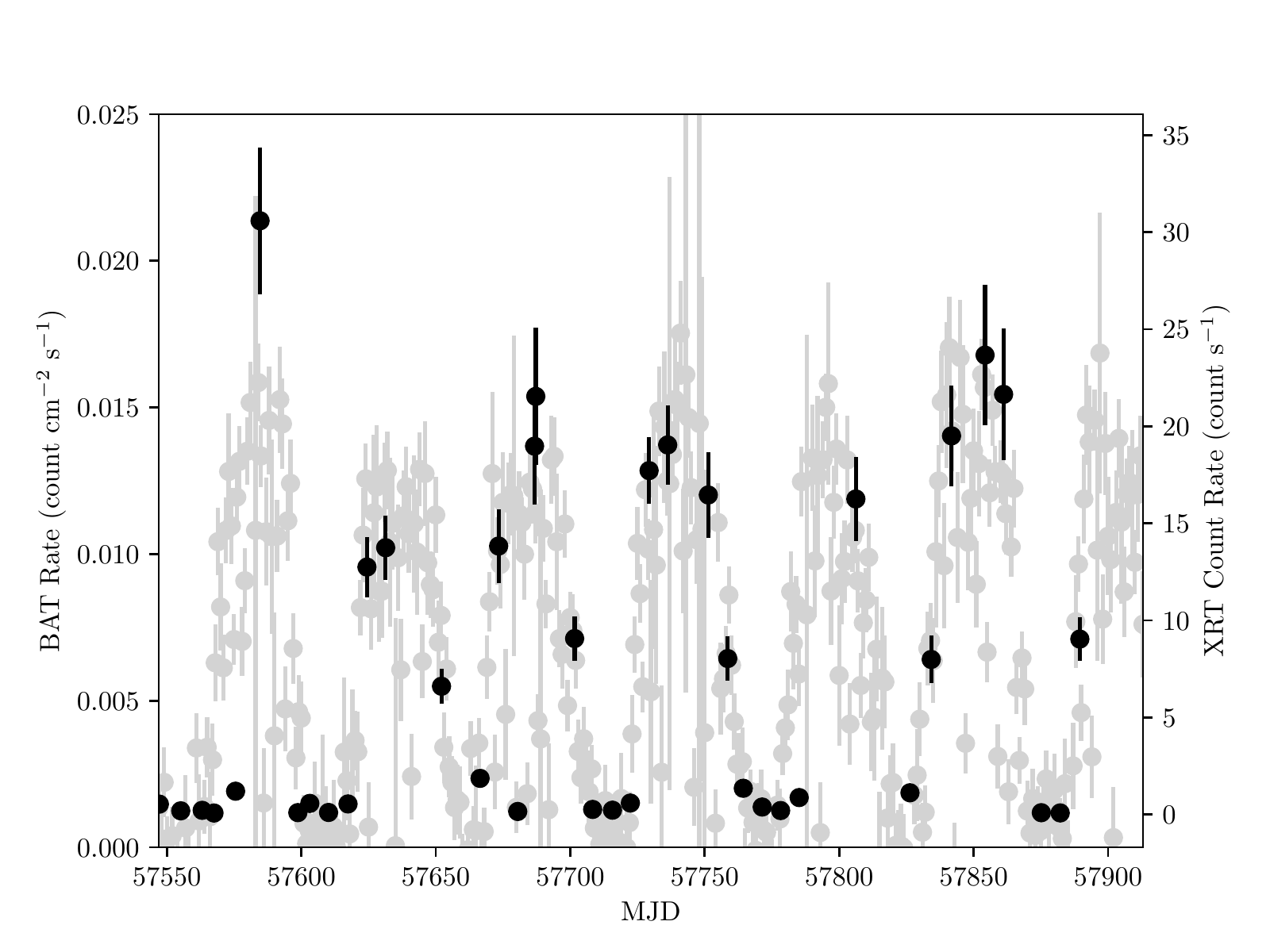}
\caption{\label{fig:smcx1batxrt}Comparison of the BAT Transient Monitor light-curve of SMC X-1 (light grey) and the \scubed{} detections (black).}
\end{figure}

The BAT Transient Monitor \citep{Krimm13} detects SMC~X-1 daily. In Figure~\ref{fig:smcx1batxrt}, we compare the count rate seen in those data with the \scubed{} derived count rate. It is clear that the \scubed{} data, although much more poorly sampled than the BAT Transient Monitor data, which are plotted here with 1~day time resolution, closely follows the super-orbital variations of SMC X-1. L-S analysis of the BAT Transient Monitor data reveals a peak in the periodogram at 53.5~days, and analysis of the sparser \scubed{} observation data shows a peak in the periodogram at 53.4~days, i.e. the periods measured by BAT and \scubed{} are consistent, as expected. 

\subsubsection{SMC X-3\label{sec:smcx3}}

\smcxthree\ is a \bexrb\ source in the SMC, first discovered by the SAS~3 satellite in 1977 \citep{Li77,Clark78}. In 2002, \rxte\ detected a bright outbursting pulsar in the SMC with a $\sim7.8$~s periodicity, although due to the localization accuracy of \rxte\ it was not possible to associate this with \smcxthree. \cite{Edge04}, utilizing \chandra\ observations, accurately localized the \rxte\ discovered pulsar, and confirmed that it was indeed a new outburst of \smcxthree. 

\smcxthree{} is known to have large \typetwo{} outbursts, however no outburst from the source had been seen during the \swift{} mission lifetime, although the source had been observed in quiescence several times by \swift{}. \replacedref{Analysis of \scubed{} observations of the region show that the source}{\smcxthree{}} was detected during the first \scubed{} observation on 2016 June 8. \scubed{} observed \smcxthree{} 5 times between 2016 June 24 and 2016 July 16 (see Table~\ref{tab:obs}), but it was not detected in any of those observations.

\smcxthree\ was detected again on 2016 July 30, $\sim5$ times brighter than the previous 2016 June 8 detection. The brightening continued, showing that the source had entered into a significant likely \typetwo{} outburst. Unfortunately at this time, due to an error in the analysis software, this brightening was not flagged, and therefore went unnoticed. However, two days previously, on 2016 August 8, \maxi{} reported the detection of a bright new transient, named MAXI~J0058$-$721, in the SMC consistent with the location of \smcxthree\ \citep{NegoroATEL9348}, but also with SXP~6.85 (SC49 in Table~\ref{tab:scubedsources}), which was undergoing an outburst at the time \citep{KenneaATel9299}.

The \scubed{} observations of 2016 August 8 confirmed that the MAXI~J0057$-$721 was in fact \smcxthree\ entering a new outburst \citep{KenneaATel9362}. \scubed{} monitoring (see Figure~\ref{fig:flaring}), tracked the outburst of \smcxthree. In addition to \scubed{} monitoring, additional Swift TOO observations were performed in WT mode in order to track the evolution of the pulsar periodicity, which not only confirmed the presence of the $\sim7.8$~s period, but also allowed the accretion-powered spin-up of the pulsar to be measured. Results of those observations are reported by \cite{Townsend2017}, and additionally by \cite{Tsygankov17}, \cite{Weng17} and \cite{Koliopanos18}. By measuring the effects of orbit induced Doppler shift on the spin period of \smcxthree{}, these data provided for the first time a dynamical measure of the orbital period, $P_\mathrm{orb} = 45.04 \pm 0.08$~days, and eccentricity $e=0.244 \pm 0.005$ (values quoted from \citealt{Townsend2017}).

Although the outburst of \smcxthree\ appears to have ended in late February 2017 (\scubed{} measured flux level returned to ~0.1 c/s, consistent to its pre-outburst level on the observation of 2017 February 22), it continued to be detected, although the detections appear interspersed with periods of non-detection. Examining combined \scubed{} and TOO observations performed by \swift{} in PC mode, which were requested as part of the additional monitoring of \smcxthree, reveals that these detections are in fact low-level periodic outbursts, peaking every $\sim45$~days, i.e. the orbital period of \smcxthree.

\begin{figure}[t]
\centering
\includegraphics[width=0.5\textwidth]{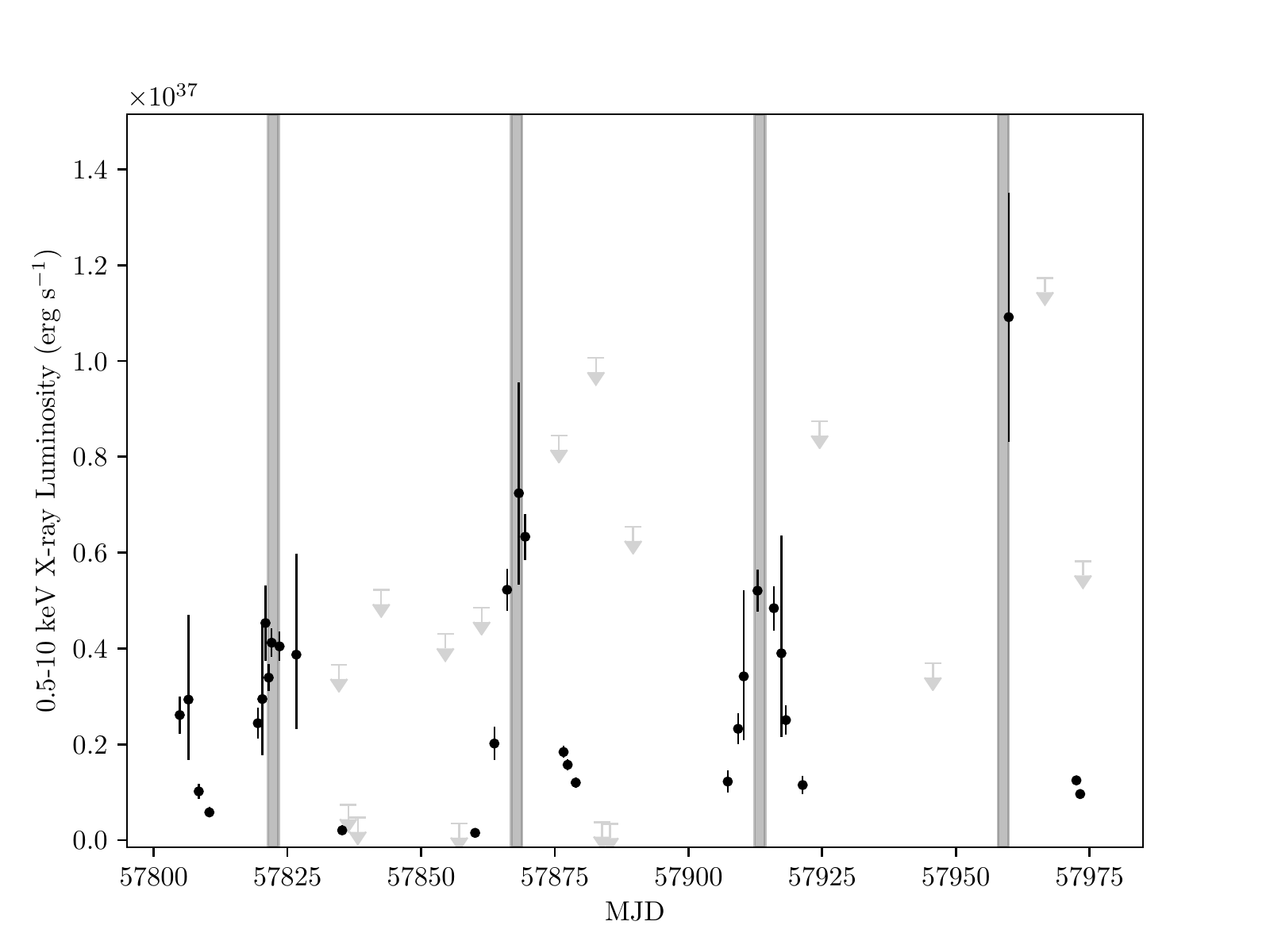}
\caption{\label{fig:smcx3type1}Post outburst monitoring of \smcxthree\ by \scubed{}, with additional Swift/XRT data taken as part of a TOO campaign. Repeated flares at the orbital period of $\sim45$~days are the signature of \typeone{} outbursts from this \bexrb\ source. Predicted periastron passages, derived from the orbital ephemeris of \cite{Townsend2017}, are shown as vertical lines.}
\end{figure}

Figure~\ref{fig:smcx3type1} shows the post-outburst light-curve of \smcxthree, with the predicted periastron passages using the orbital ephemeris reported by \cite{Townsend2017} shown. These periodic flares therefore represent the signature of \bexrb\ \typeone{} outbursts from \smcxthree, peaking at the at the orbital periastron. These \typeone{} bursts have been reported at a similar level in pre-outburst \swift{} data by \cite{Tsygankov17}, and therefore show no evidence of enhancement related to the large \typetwo{} outburst.
Figure~\ref{fig:smcx3type1} shows the post-outburst \typeone{} bursts, which peak between $\sim4 \times 10^{36}$ to $\sim10^{37}~\mathrm{erg s^{-1}}$ (0.5 -- 10 keV), consistent with the typical range of outburst luminosities seen in \typeone{} outbursts from \bexrbs{} (e.g.~\citealt{Stella86}).

\subsubsection{SXP~59.0}

SXP~59.0 (\onex{}~J005456.4-722647, \scubed{} source SC403) is an \bexrb\ system, first identified as an X-ray pulsar in outburst by \rxte\ in observations taken on 1998 January 20, with a measured period of $59.0 \pm 0.2$~s and designated XTE J0055-724 \citep{MarshallIAUC6818}. SXP~59.0 has a reported orbital period of $122.1 \pm 0.38$~days \citep{Galache08} based on \rxte{} observations, and an independently derived optical derived period of $122.25$~days, consistent with the \rxte\ period, was reported from analysis of OGLE I-band light curves by \cite{Bird12}.

\begin{figure}[t]
\centering
\includegraphics[width=0.5\textwidth]{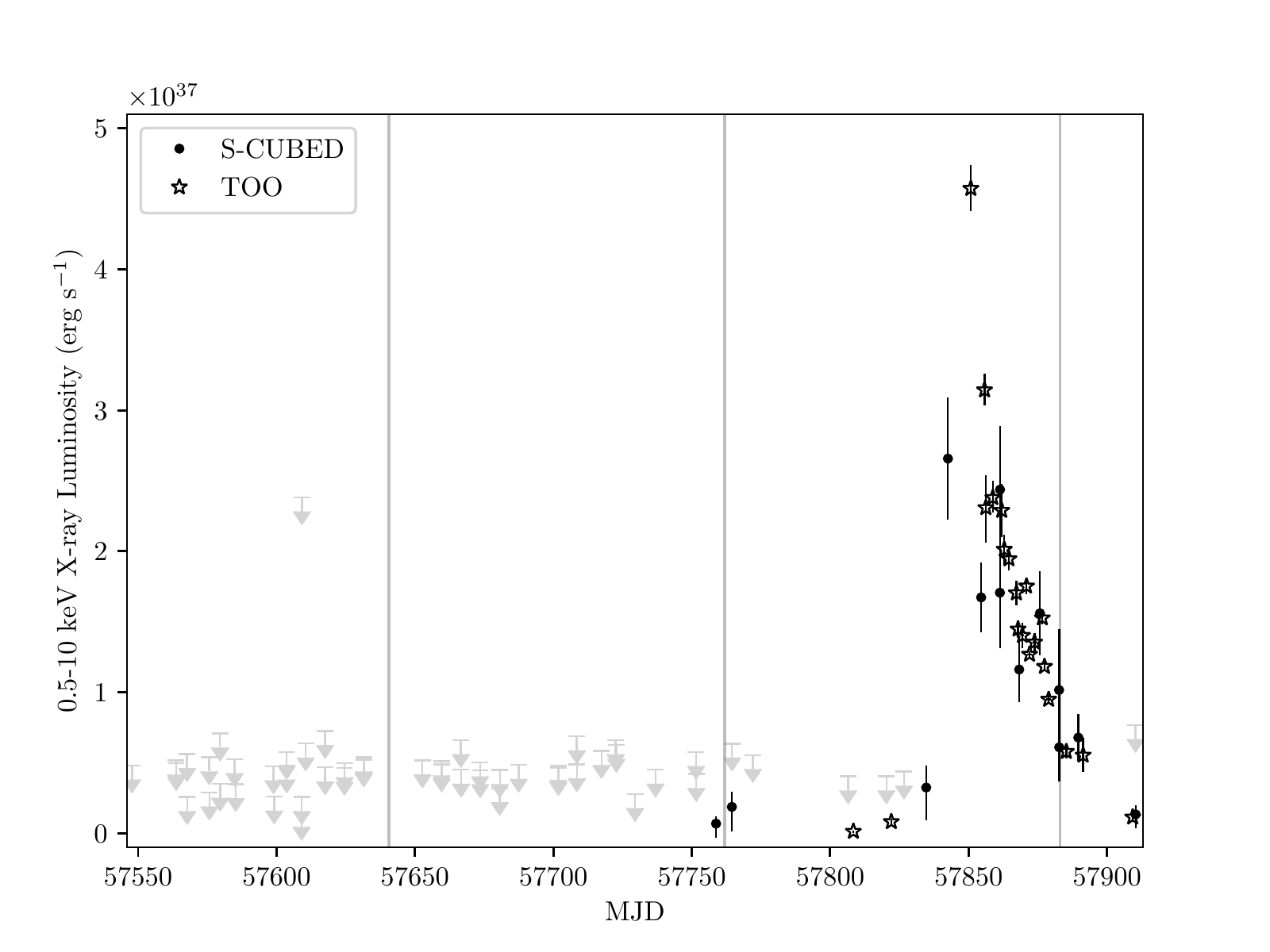}
\caption{\label{fig:sxp59}Light-curve of combined \scubed{} (circles) and Swift TOO observations (stars) of SXP~59.0, which showed a bright \typetwo{} outburst starting around 2017 March 30. The source appears to have returned to the pre-outburst observation level at the time of the final \scubed{} observation on 2017 Jun 6.}
\end{figure}

\scubed{} detected an outburst of SXP~59.0 starting on 2017 March 30 \citep{KenneaATel10250}. As a result of this outburst, \swift{} TOO observations were requested to perform higher cadence and more sensitive monitoring of the outburst. WT mode was requested in order to avoid pile-up, and also to help with detection of the pulsar period. WT mode observations were taken between 2017 April 12 and 2017 April 30, every 3 days, with a requested observation time of 3ks per observation, although the actual exposure times varied due to scheduling issues. In addition to these WT observations, several deep exposures in PC mode were taken during this period in coordination with \nustar{} observations of the SMC and SXP~59.0. \scubed{} observations continued to be taken during this period also. The combined light-curve of \scubed{} and TOO (including serendipitous) observations are shown in Figure~\ref{fig:sxp59}. 

The outburst was observed to peak during a TOO observation on 2017 April 07 at $L_\mathrm{X} = 4.573^{+0.159}_{-0.168} \times 10^{37}~\mathrm{erg~s^{-1}}$, approximately $26\%~L_\mathrm{Edd}$ for a $1.4~M_\odot$ NS. After peak, the outburst declines approximately exponentially with an time constant of $\tau \simeq 15.9$ days. Given the brightness and length of the outburst, this was likely a \typetwo{} burst.

\begin{figure}[t]
\centering
\includegraphics[width=0.5\textwidth]{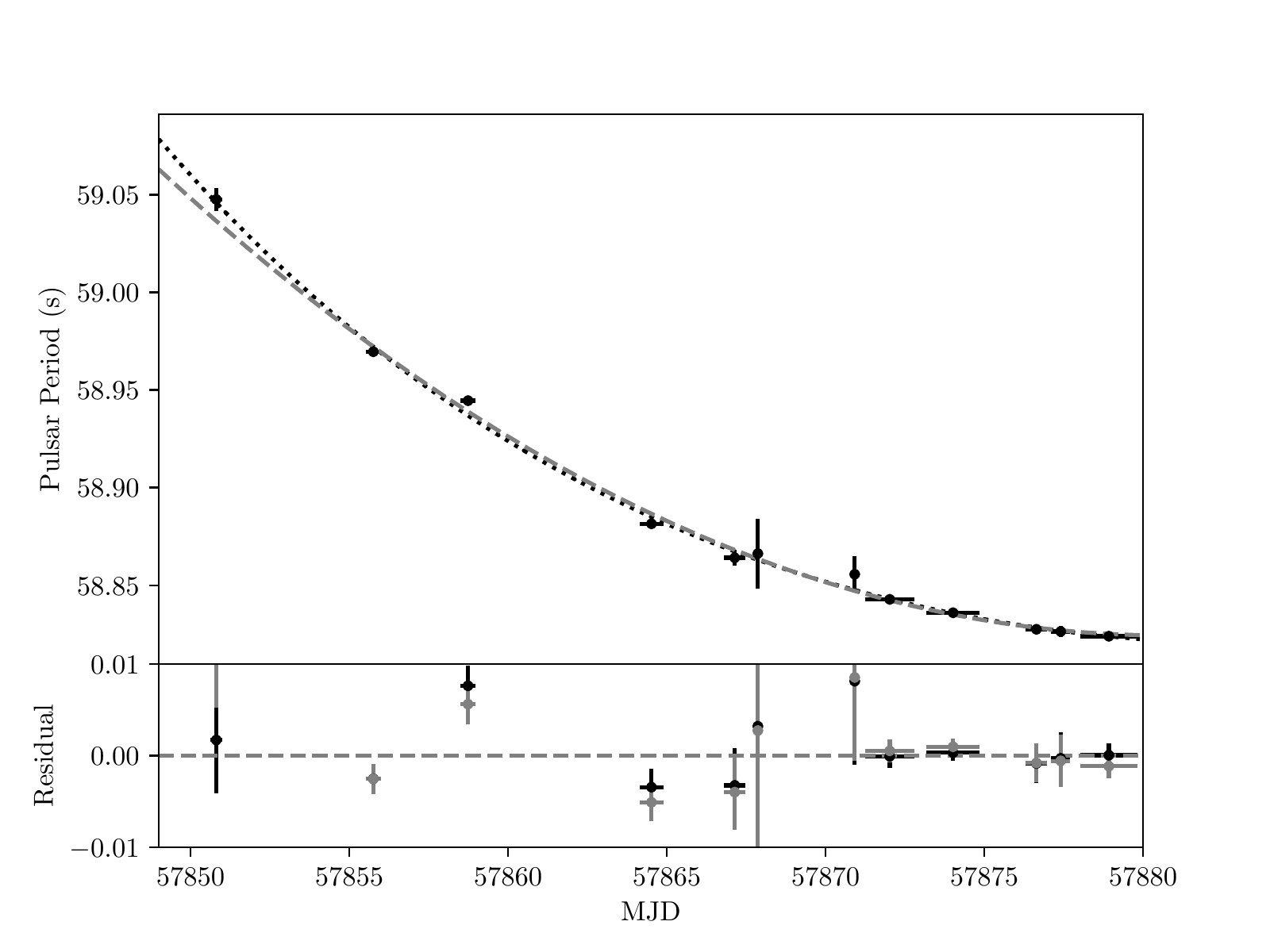} 
\caption{\label{fig:sxp59period}The pulsar period evolution of SXP~59.0 during its outburst in early 2017, utilizing WT data taken as part of a TOO campaign to study the outburst. Two models are fit to the data: a simple spin-up with $\dot{P}$ and $\ddot{P}$ components (black dotted line), and a Doppler-shifted pulsar spin at the orbital period of 122.25~days (grey dashed line). }
\end{figure}

In the \swift{} observation of 2017 April 7, a pulsar period of $P = 59.0476 \pm 0.00165$~s was detected. Measurements of the pulsar period are shown in Figure~\ref{fig:sxp59period}, after 2017 May 5, SXP~59.0 became to faint for an accurate period to be detected. The pulsar period during this time shows significant time variability. Modeling this as a simple spin-up, we obtain a $\dot{P} = -1.69 \pm 0.07 \times 10^{-7}$~s/s, and $\ddot{P} = -5.49 \pm 0.40 \times 10^{-9}$~s/s/s. However, in these systems, pulsar spin period is often significantly affected by orbital motion induced Doppler shift, for example as seen in \swift{} observations of \smcxthree\ \citep{Townsend2017} and SXP~5.05 \citep{Coe15}, so we cannot rule out that orbital motion contributes to the observed pulsar period changes.

In order to estimate the probability of this, we fit a model consisting of a simple spin-up modified by Doppler shifted orbital motion, similar to the method employed by \cite{Coe15}. As the orbital period is much longer than the $\sim24$~days in which the pulsar period was detected, we fixed the orbital period to 122.25~days, and fixed the orbital eccentricity to typical value of $e=0.3$. The resultant fit is improved over the spin-up only fit, reduced $\chi^2 = 2.13$ (9 dof) versus $\chi^2 = 2.57$ (9 dof) for the spin-up only fit, however derived orbital parameters are not well constrained. For the fit with orbital modulation, we derive an underlying spin-up $\dot{P} = -5.49 \times 10^{-9}$~s/s, although given the uncertainties on the orbital modeling, we consider this value to have likely larger uncertainties than the quoted fit errors.

It is a clear that in order to derive an orbital solution for SXP~59.0 utilizing pulsar timing, a longer outburst in which the spin-period of the pulsar was measurable for at least one full orbital period would be necessary. \editpre{We note that an L-S search of the light-curve of SXP~59.0 (see Section~\ref{sec:period_searches}), did not find a significant period. SXP~59.0 does show possible evidence of a single \typeone{} outburst, however the \scubed{} light-curve is dominated by the \typetwo{} outburst during the period when a 2nd periastron passage would occur, making detection of the relatively long (122~day) orbital period not possible.}

\subsubsection{SXP~6.85}\label{sec:sxp685}

SXP~6.85 (SC49) is an \bexrb{} in the SMC, first found by \rxte{} and named XTE J0103-728 \citep{ATel163}. It is associated with a Be-star companion based on localization by \xmm{} during an outburst seen in 2006 October \citep{Atel1095}, and is therefore a \bexrb{}. Analysis of OGLE data found an likely orbital period of $P_\mathrm{orb} \simeq 24.8$~days \citep{Atel7498}. A further outburst of this system was detected by \integral{} on 2015 April \citep{Atel7481}.

The \scubed{} light-curve of SXP~6.85 is shown in Figure~\ref{fig:flaring}. Initial observations  did not detect the source, however starting 2016 July 29, SXP~6.85 was significantly detected, with peak luminosities of $1.6\pm0.4 \times 10^{37}\ \mathrm{erg~s^{-1}}$ (0.5 - 10 keV at 62 kpc). 
Examination of archival data at this location reveals that \swift\ observed SXP~6.85 serendipitously in PC mode for a total of 15.7~ks between 2016 March 25 and 2016 March 28, as it lay near the GRB~160325A \citep{GCN19222}. During that observation, no photons were detected from the location of SXP~6.85, allowing us to place strong upper limits on the X-ray count rate of the source approximately 4 months before its 2016 July 29 outburst of $<2.4 \times 10^{34}\ \mathrm{erg\ s^{-1}}$. No follow-up observations were made to confirm the pulsar period, however the \scubed{} localization is consistent with that reported by \xmm{}.

After the initial outburst detection, SXP~6.85 appears to fade over a period of ~50 days (see Figure~\ref{fig:flaring}, given the purported 24.8~day orbital period of this \bexrb{}, we conclude that this is a \typetwo{} outburst. Starting 2016 December 28, the source appears to undergo a second flare, peaking around 2017 February 2017.

As reported in Table~\ref{tab:periods}, SXP~6.85 displays a large discrepancy between the orbital period of $\sim24.8$~days and the \scubed{} detected period of $161.6$~days. Given the presence of two similar level, likely \typetwo{}, flares in the \scubed{} light-curve, the origin of this detected period can be explained by it being the time between these two events. As only two full cycles at this period exist in the first year of \scubed{} data, more observations over a longer timescale are needed to confirm if these outbursts are indeed periodic. 

\subsubsection{SXP~2.16}

SXP~2.16 is a \bexrb{}, first discovered by \rxte{} (as XTE~J0119$-$731; \citealt{CorbetIAUC8064}). In 2014 a bright transient in the SMC, IGR~J01217$-$7257, was identified in outburst in by \integral, however a follow-up \swift{} WT mode observation could not confirmed the presence of pulsations to confirm that it was indeed SXP~2.16 \citep{CoeATEL5806}. An \xmm\ observation performed in 2015, detected the presence of the 2.16s periodicity, confirming that IGR~J01217$-$7257 and XTE~J0119$-$731 are the same source: SXP~2.16 \citep{Vasilopoulos17b}.

Some results from the \scubed{} observations of SXP~2.16 have been previously reported by \cite{Boon17}. For the majority of the period of interest reported in this paper, SXP~2.16 was not detected. However, it underwent two short outbursts in which it was above the detection limit. SXP~2.16 was detected in three consecutive \scubed{} observations that took place on 2016 July 6, 2016 July 10, and 2016 July 15, as reported by \cite{Boon17}. Additional detections of SXP~2.16 were made by \scubed{} during observations taking place on 2016 September 28 and 2016 October 5. \cite{Boon17} reports estimates of the orbital period from BAT data and OGLE I-band light-curves of $82.5\pm0.7$~days and $83.67\pm0.05$~days respectively. The time between the onset of the two periods of detection in \scubed{} is 84~days, which is consistent with these outbursts recurring at the reported orbital period, suggesting that these are likely \typeone{} outbursts. 

\begin{figure}[t]
\centering
\includegraphics[width=0.5\textwidth]{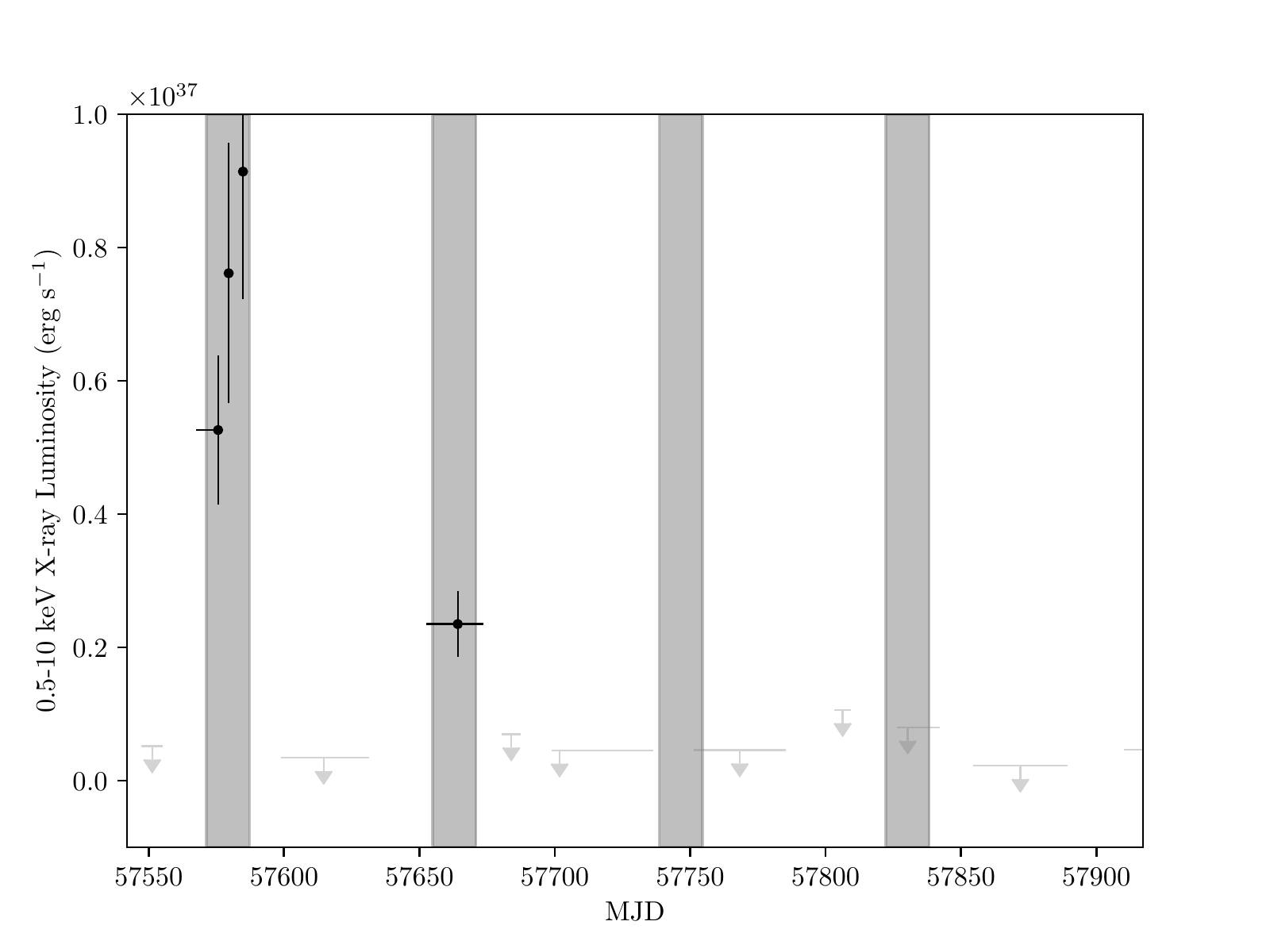}
\caption{\label{fig:sxp216type1}\scubed{} light curve of SXP~2.16. Predicted periastron passages from at an orbital period of 83~days, based on the orbital period reported by \cite{Boon17} are marked as vertical gray regions (width set to that of the first detected outburst). Enhanced emission during two of the predicted periastron passages are seen, but no similar outbursts are seen for the next three predicted periastron passages. However, we cannot rule out that these outbursts were simply missed due to the low sampling rate of \scubed{}.}
\end{figure}

The \scubed{} light-curve of SXP~2.16 is shown in Figure~\ref{fig:sxp216type1}, with gray lines representing estimates of the expected periastron passage times, assuming an orbital period of $P_{\mathrm{orb}} = 82.5$~days, and assuming a periastron passage epoch of 2016 June 21, based on the center of the first outburst seen by \scubed{}. 

It should be noted that no other outbursts were seen, despite two more periastron passages occurring during the period of interest. Although the lengths of the detected outbursts are uncertain due to the relatively infrequent sampling of \scubed{}, the time between the first and last detection of the first outburst was 9.2~days, and for the second outburst was 7~days, we therefore suggest that 7~days should be considered the lower-limit for the outburst timescale.

\scubed{} observations were performed on 2016 December 14 (10~days before) the predicted periastron passage centered around 2016 December 24, and 2016 December 28 (4~days after), given this it is plausible that the expected \typeone{} outburst was simply missed due to the fact that \scubed{} observations were taken 15 days apart during this period.

For the final periastron passage, centered around 2017 March 17, observations by \scubed{} were taken on 2017 March 14 (3 days before) and 2017 March 22 (5 days later), an 8 day gap. It is clear that an outburst lasting $>10$~days would likely have been detected during this observation period. However, we cannot strongly rule out having missed an $<8$~day long outburst. During the reported period, no additional observations of SXP~2.16 were taken by \swift{}.

A LS periodogram search of the \scubed{} light-curve reveals a peak periodicity at $80.25 \pm 0.40$~days as reported in Table~\ref{tab:periods}, which is somewhat inconsistent with the values reported by \cite{Boon17}. However, an L-S search of the only first 0.5 years of \scubed{} data, which contains the two outbursts, finds a peak period of $83.18 \pm 0.45$~days, consistent within errors with the BAT derived orbital period, and close to the orbital period.

\subsubsection{SXP~15.6}\label{sec:SXP15.6}

SXP~15.6 (SC3) was detected by \scubed{} in an observation taken on 2016 June 16, as a X-ray point source with a position consistent with location of the cataloged X-ray source XMMU~J004855.5$-$734946, reported to be a possible HMXB \citep{Haberl16}. In the following observation taken on 2016 June 24, it was found to have brightened, \deleted{significantly} reaching a luminosity of $\sim8 \times 10^{37}\ \mathrm{erg\ cm^{-2}}$, significantly brighter than the cataloged \xmm{} brightness of $6.4 \times 10^{-13}\ \mathrm{erg\ cm^{-2}\ s^{-1}}$ \citep{EvansATEL9197}, which is at 62~kpc is a luminosity of $\sim3 \times 10^{35}\ \mathrm{erg\ s^{-1}}$. \xmm\ observations revealed a $\sim15.6$~s period, strongly suggesting that the source is a \bexrb{} \citep{Vasilopoulos17a}. SXP~15.6 was reported to have an optical counterpart in the catalog of \cite{Evans04}, with a spectral type of B0~IV-Ve, firmly confirming it as a \bexrb{} \citep{McBride17}.

The \scubed{} light-curve of SXP~15.6 shows brief flaring intervals (Figure~\ref{fig:flaring}), along side periods of non-detection, which appear to be periodic. As reported in Table~\ref{tab:periods}, SXP~15.6 is found to have a high significance period detection at $36.82\pm1.53$~days (see Table~\ref{tab:periods}). This periodicity is consistent with the orbital period derived from OGLE~IV light-curve of $36.43 \pm 0.01$~days \citep{McBride17}. We therefore suggest that the \scubed{} light-curve shows strong signature of repeated \typeone{} outbursts at the orbital period.

\subsubsection{SXP~202A}

SXP~202A (SC20) is a \bexrb\ first detected in observations by \xmm\ in 2003 \citep{Majid04}. \scubed{} detected an apparent outburst of SXP~202A observations taken on 2016 July 6, 2016 July 10, and 2016 July 16 \citep{CoeATEL9307}. 
This outburst peaked at a luminosity of $8.3 \pm 2.6 \times 10^{36}~\mathrm{erg~s^{-1}}$ (0.5--10~keV), consistent with the range seen for \typeone{} \bexrb\ outbursts. Outside of this outburst SXP~202A is detected in $40\%$ of \scubed{} observations at a mean luminosity of $2.8 \pm 0.6 \times 10^{36}~\mathrm{erg~s^{-1}}$ (0.5--10~keV), with upper limits calculated for the non-detections are consistent with this emission level. No other outburst was detected during the first year of \scubed{} observations. 

If the outburst seen by SXP~202A was in fact a \typeone{} outburst, no recurrence was seen in the first year of observations, although we cannot strongly rule out that a \typeone{} outburst did not occur during an observation gap. Currently SXP~202A does not have a measured orbital period, although based on the empirically measured relationship between pulsar spin and orbital period in \bexrbs{} (the ``Corbet Diagram''; \citealt{Corbet84}), we would expect the orbital period to be long ($\sim200$ days). The non-detection of repeating outbursts in SXP~202A does not allow us to place constraints on the orbital period in this system. Examination of the L-S periodogram does not show any significant peaks. Longer observations of SXP~202A will be required in order to pin down any orbital modulation.

\subsubsection{SXP~46.6}

SXP~46.6 (SC271) is a \bexrb{} \citep{CoeKirk15} with a reported orbital period of $P_\mathrm{orb} = 137.4 \pm 0.2$~days based on OGLE light-curves \citep{Bird12}, and a consistent X-ray derived $P_\mathrm{orb} = 137.36$ \citep{Galache08}.  \editpre{As noted in Section~\ref{sec:period_searches}, \scubed{} data show a significantly detected period of $143.29 \pm 4.5$ days, close to the previously reported orbital periods.}

\begin{figure}[t]
\centering
\includegraphics[width=0.5\textwidth]{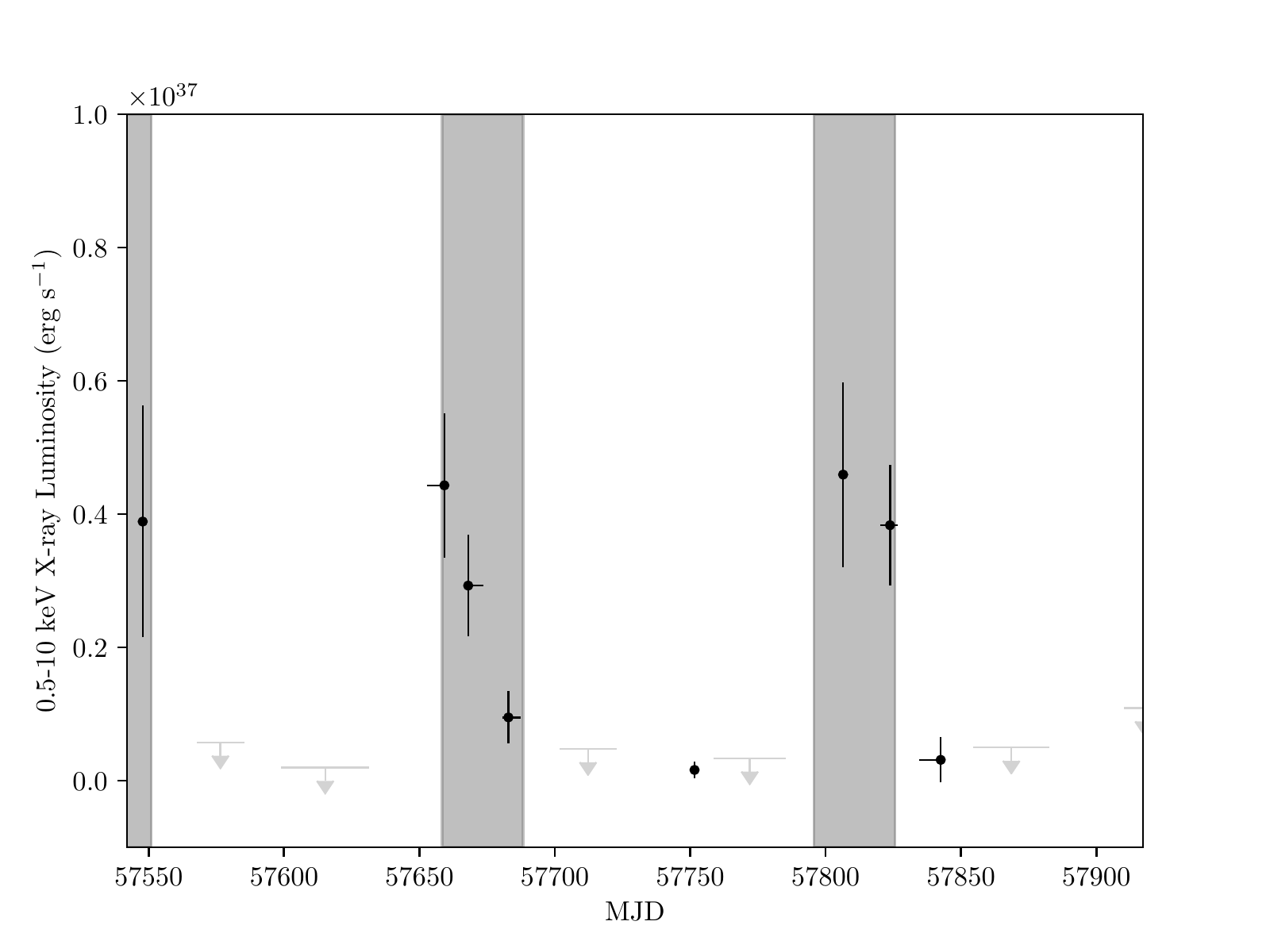}
\caption{\scubed{} light-curve of SXP~46.6 with predicted periastron passages marked in grey, based on the orbital period of 137.4~days, with the periastron passage fixed to the maximum flux point at 2016 October 19 (MJD 57670). Three orbital periods are covered by the \scubed{} monitoring, and detections of SXP~46.6 during these periods are consistent with \typeone{} outbursts.}
\label{fig:sxp466type1}
\end{figure}

The \scubed{} light-curve of SXP~46.6 (see Figure~\ref{fig:sxp466type1}) reveals apparent low-level activity from SXP~46.6 during the year of observations, with detections of the source made during three periods, firstly during the first observation on 2016 June 8, secondly during an apparent small outburst, starting 2016 September 28, and fading away with the last detection on 2016 October 19, and a final period with three detections between 2017 February 22 and 2017 March 14. Figure~\ref{fig:sxp466type1} shows the predicted period of time in which a \typeone{} outburst would occur, assuming a 137.4~day orbital period, with the 
prominent outburst in the \scubed{} data as the assumed periastron passage time. It is clear that both the early initial detection, and the latter detections are consistent with being part of a \typeone{} outburst.

\subsubsection{SXP~91.1}

SXP~91.1 (SC6) is a \bexrb{}, which was first discovered by \rxte{} (named RX J0051.3$-$7216) with a $92 \pm 1.5$~s pulsar period. \cite{Townsend13} determined an orbital period measurement of $88.42 \pm 0.14$~days, and showed that SXP~91.1 has significant trend of long-term pulsar spin-up with $\dot{P} = 1.442 \pm 0.005 \times10^{-8}\ \mathrm{s\ s^{-1}}$. Further orbital period estimates from optical data reported to be $88.2$~days \citep{Schmidtke06} and $88.37\pm0.03$~days \citep{Bird12}. 

The light-curve of SXP~91.1 shown in Figure~\ref{fig:flaring} shows periods of apparent enhanced X-ray emission, and periods where the source was not active. We detect a periodicity in \scubed{} data of $P=89.25\pm0.03$~days, consistent with the previously reported orbital period. The false alarm probability of this peak is found to be $\sim2 \times 10^{-7}$, suggesting an unambiguous detection of the orbital period in the \scubed{} data. 

\scubed{} monitoring shows two well detected \typeone{} outbursts centered on 2016 December 8 and 2017 March 6. Three other periastron passages show enhanced X-ray emission, but are not as well sampled, and the peak emission were likely missed. 

\section{Discussion and conclusions}

This paper presents results from the first year of the \scubed{} survey of the SMC. \scubed{} is focused on discovery of new X-ray outbursts by known and unknown X-ray transients, which in the case of the SMC are mostly \bexrbs{} where the X-ray emission is due to \typeone{} and \typetwo{} X-ray outbursts. 

During the first year of \scubed{} observations, two major \typetwo{} outbursts of \bexrbs{} were detected in the SMC: \smcxthree{} and SXP~59.0. In \editpre{addition to}\deleted{two} these two bright outbursts, SXP~6.85 showed strong evidence two extended outbursts, that are likely also \typetwo{} in nature. The 2016--2017 super-Eddington (peaking at $1.4\times10^{39}\ \mathrm{erg\ s^{-1}}$) \typetwo{} outburst of \smcxthree{}, which was first identified by \scubed{}, and has been reported on extensively in literature \citep{Townsend2017,Tsygankov17,Weng17}, shows the power of the \scubed{} observing technique to catch bright outbursts early. 
\scubed{} triggered extensive follow-up campaigns, including \swift{} deeper observations, which were able to measure the spin evolution of the pulsar during the bright \typeone{} outburst and model the orbital parameters. 

The bright \typetwo{} outburst from SXP~59.0, also triggered follow-up observations with Swift. The peak flux of the outburst of SXP~59.0 was much fainter than the outburst of \smcxthree{}, however this outburst was quickly identified by \scubed{}. Results from both \scubed{} and \swift{} TOO observations of SXP~59.0 are reported for the first time in this paper. 

In addition to the bright \typetwo{} outbursts, direct evidence of \typeone{} outbursts were seen from at least three known \bexrbs{}: \smcxthree{} after the \typetwo{} outburst had ended; SXP~46.6, which was mostly not detected except for for three outburst which occurred at intervals consistent with the reported 137.4~day orbital period; and SXP~2.16 which showed two outbursts separated by $\sim83$ days, close to the reported orbital period, but no more obvious outbursts after that. 

We have performed period searches of the \scubed{} data in order to search for any possible orbital periods. \scubed{} detected emission from a total of 29 SXP sources for which the periods have been previously reported either from optical or X-ray measurements. Of those 29 sources we positively detect an orbital period for 8 of them\deleted{those objects} (see Table~\ref{tab:periods}). 

Observing the bulk properties of sources in the \scubed{} survey, we measured the DC, and note that the DC plot in Figure~\ref{fig:DCvsOrb} does show evidence that the maximum value of DC decreases with orbital period. This raises the question, are we missing \typeone{} outbursts? In many cases it can be explained that \scubed{} is simply missing them due to the non-uniform nature of the coverage. However, the most likely scenario is that the outbursts are below the level of detection for these short exposures. For example, \scubed{} monitoring has shown that in the case of objects like SXP~2.16, that the flux level of \typeone{} outbursts varies. \cite{Reig11} has shown examples of transient and non-transient behaviors of \typeone{} bursts in \bexrbs{}: EXO~2030+375, which shows regular \typeone{} bursts; and 4U~0115+63 which shows only occasional detections of \typeone{} bursts. Given this, the wide scatter in the measured DC and the presence of an apparent maximum DC which decreases as orbital period gets larger, is the expected behavior.

Results from the first year of \scubed{} presented here show both the strengths and weaknesses of the survey. For detection of \typeone{} outbursts, comparing the results from the post-outburst TOO observations of \smcxthree{} to \scubed-only data, it would clearly be beneficial to both observe more frequently, and with a longer exposure in order to better detect these events. However, due to limitations in \editpre{the amount of} available observing time, it is not likely that this could be achieved with \swift{} without significantly changing the design of the survey. However, where periods were detected in the \scubed{} survey, they have very good agreement with those previously reported in the literature.

\editpre{Out of 265 \scubed{} detected `Good' X-ray sources in the vicinity of the SMC during the first year of the survey, the majority (160) are previously uncataloged. Many of these sources are in regions previously surveyed to much higher sensitivities by \chandra{}, \xmm{} and others. We believe that these sources are real to high confidence, and represent the discovery of a large population of variable sources, which could have only be discovered by performing repeated searches, rather than a single deep survey. This result strongly validates the approach of \scubed{}.}

\section{Acknowledgements}

JAK acknowledges the support of NASA contract NAS5-00136, and NASA grant NNX16AR15G (through the Swift Guest Investigator program). 
This work made use of data supplied by the UK Swift Science Data Centre at the University of Leicester.
The Digitized Sky Surveys were produced at the Space Telescope Science Institute under U.S. Government grant NAG W-2166. The images of these surveys are based on photographic data obtained using the Oschin Schmidt Telescope on Palomar Mountain and the UK Schmidt Telescope. The plates were processed into the present compressed digital form with the permission of these institutions.

\clearpage
\bibliographystyle{aasjournal}
\bibliography{scubed}

\begin{thebibliography}{}
\expandafter\ifx\csname natexlab\endcsname\relax\def\natexlab#1{#1}\fi
\providecommand{\url}[1]{\href{#1}{#1}}

\bibitem[{{Antoniou} {et~al.}(2009){Antoniou}, {Zezas}, {Hatzidimitriou}, \&
  {McDowell}}]{Antoniou09}
{Antoniou}, V., {Zezas}, A., {Hatzidimitriou}, D., \& {McDowell}, J.~C. 2009,
  \apj, 697, 1695

\bibitem[{{Astropy Collaboration} {et~al.}(2013){Astropy Collaboration},
  {Robitaille}, {Tollerud}, {Greenfield}, {Droettboom}, {Bray}, {Aldcroft},
  {Davis}, {Ginsburg}, {Price-Whelan}, {Kerzendorf}, {Conley}, {Crighton},
  {Barbary}, {Muna}, {Ferguson}, {Grollier}, {Parikh}, {Nair}, {Unther},
  {Deil}, {Woillez}, {Conseil}, {Kramer}, {Turner}, {Singer}, {Fox}, {Weaver},
  {Zabalza}, {Edwards}, {Azalee Bostroem}, {Burke}, {Casey}, {Crawford},
  {Dencheva}, {Ely}, {Jenness}, {Labrie}, {Lim}, {Pierfederici}, {Pontzen},
  {Ptak}, {Refsdal}, {Servillat}, \& {Streicher}}]{Astropy}
{Astropy Collaboration}, {Robitaille}, T.~P., {Tollerud}, E.~J., {et~al.} 2013,
  \aap, 558, A33

\bibitem[{{Azzopardi} \& {Vigneau}(1979)}]{Azzopardi79}
{Azzopardi}, M., \& {Vigneau}, J. 1979, \aaps, 35, 353

\bibitem[{{Barthelmy} {et~al.}(2005){Barthelmy}, {Barbier}, {Cummings},
  {Fenimore}, {Gehrels}, {Hullinger}, {Krimm}, {Markwardt}, {Palmer},
  {Parsons}, {Sato}, {Suzuki}, {Takahashi}, {Tashiro}, \&
  {Tueller}}]{Barthelmy05}
{Barthelmy}, S.~D., {Barbier}, L.~M., {Cummings}, J.~R., {et~al.} 2005, \ssr,
  120, 143

\bibitem[{{Bird} {et~al.}(2012){Bird}, {Coe}, {McBride}, \& {Udalski}}]{Bird12}
{Bird}, A.~J., {Coe}, M.~J., {McBride}, V.~A., \& {Udalski}, A. 2012, \mnras,
  423, 3663

\bibitem[{{Boon} {et~al.}(2017){Boon}, {Bird}, {Coe}, {Corbet}, {Evans},
  {Kennea}, {Krimm}, {Laycock}, \& {Udalski}}]{Boon17}
{Boon}, C.~M., {Bird}, A.~J., {Coe}, M.~J., {et~al.} 2017, \mnras, 466, 1149

\bibitem[{{Bradt} {et~al.}(1993){Bradt}, {Rothschild}, \& {Swank}}]{Bradt93}
{Bradt}, H.~V., {Rothschild}, R.~E., \& {Swank}, J.~H. 1993, \aaps, 97, 355

\bibitem[{{Burrows} {et~al.}(2005){Burrows}, {Hill}, {Nousek}, {Kennea},
  {Wells}, {Osborne}, {Abbey}, {Beardmore}, {Mukerjee}, {Short}, {Chincarini},
  {Campana}, {Citterio}, {Moretti}, {Pagani}, {Tagliaferri}, {Giommi},
  {Capalbi}, {Tamburelli}, {Angelini}, {Cusumano}, {Br{\"a}uninger}, {Burkert},
  \& {Hartner}}]{Burrows05}
{Burrows}, D.~N., {Hill}, J.~E., {Nousek}, J.~A., {et~al.} 2005, \ssr, 120, 165

\bibitem[{{Carpano} {et~al.}(2017){Carpano}, {Haberl}, \&
  {Sturm}}]{Carpano2017}
{Carpano}, S., {Haberl}, F., \& {Sturm}, R. 2017, \aap, 602, A81

\bibitem[{{Casares} {et~al.}(2014){Casares}, {Negueruela}, {Rib{\'o}}, {Ribas},
  {Paredes}, {Herrero}, \& {Sim{\'o}n-D{\'{\i}}az}}]{Casares14}
{Casares}, J., {Negueruela}, I., {Rib{\'o}}, M., {et~al.} 2014, \nat, 505, 378

\bibitem[{{Cioni} {et~al.}(2013){Cioni}, {Kamath}, {Rubele}, {van Loon},
  {Wood}, {Emerson}, {Gibson}, {Groenewegen}, {Ivanov}, {Miszalski}, \&
  {Ripepi}}]{Cioni13}
{Cioni}, M. R.~L., {Kamath}, D., {Rubele}, S., {et~al.} 2013, \aap, 549, A29

\bibitem[{{Clark} {et~al.}(1978){Clark}, {Doxsey}, {Li}, {Jernigan}, \& {van
  Paradijs}}]{Clark78}
{Clark}, G., {Doxsey}, R., {Li}, F., {Jernigan}, J.~G., \& {van Paradijs}, J.
  1978, \apjl, 221, L37

\bibitem[{{Coe} {et~al.}(2015){Coe}, {Bartlett}, {Bird}, {Haberl}, {Kennea},
  {McBride}, {Townsend}, \& {Udalski}}]{Coe15}
{Coe}, M.~J., {Bartlett}, E.~S., {Bird}, A.~J., {et~al.} 2015, \mnras, 447,
  2387

\bibitem[{{Coe} {et~al.}(2014){Coe}, {Bird}, {McBride}, {Bartlett}, {Townsend},
  {Haberl}, {Kennea}, \& {Udalski}}]{CoeATEL5806}
{Coe}, M.~J., {Bird}, A.~J., {McBride}, V.~A., {et~al.} 2014, The Astronomer's
  Telegram, 5806

\bibitem[{{Coe} {et~al.}(2005){Coe}, {Edge}, {Galache}, \& {McBride}}]{Coe05}
{Coe}, M.~J., {Edge}, W.~R.~T., {Galache}, J.~L., \& {McBride}, V.~A. 2005,
  \mnras, 356, 502

\bibitem[{{Coe} {et~al.}(2016{\natexlab{a}}){Coe}, {Evans}, {Kennea}, \&
  {Udalski}}]{CoeATEL9307}
{Coe}, M.~J., {Evans}, P.~A., {Kennea}, J.~A., \& {Udalski}, A.
  2016{\natexlab{a}}, The Astronomer's Telegram, 9307

\bibitem[{{Coe} {et~al.}(2016{\natexlab{b}}){Coe}, {Kennea}, {Evans},
  {McHardy}, \& {Udalski}}]{CoeATEL9414}
{Coe}, M.~J., {Kennea}, J.~A., {Evans}, P.~A., {McHardy}, I., \& {Udalski}, A.
  2016{\natexlab{b}}, The Astronomer's Telegram, 9414

\bibitem[{{Coe} {et~al.}(2016{\natexlab{c}}){Coe}, {Kennea}, {Townsend},
  {McBride}, {Evans}, \& {Udalski}}]{KenneaATel9677}
{Coe}, M.~J., {Kennea}, J.~A., {Townsend}, L.~J., {et~al.} 2016{\natexlab{c}},
  The Astronomer's Telegram, 9677

\bibitem[{{Coe} \& {Kirk}(2015)}]{CoeKirk15}
{Coe}, M.~J., \& {Kirk}, J. 2015, \mnras, 452, 969

\bibitem[{{Coe} {et~al.}(2010){Coe}, {Bird}, {Buckley}, {Corbet}, {Dean},
  {Finger}, {Galache}, {Haberl}, {McBride}, {Negueruela}, {Schurch},
  {Townsend}, {Udalski}, {Wilms}, \& {Zezas}}]{Coe10}
{Coe}, M.~J., {Bird}, A.~J., {Buckley}, D.~A.~H., {et~al.} 2010, \mnras, 406,
  2533

\bibitem[{{Corbet} {et~al.}(2003{\natexlab{a}}){Corbet}, {Markwardt},
  {Marshall}, {Coe}, {Edge}, \& {Laycock}}]{CorbetIAUC8064}
{Corbet}, R., {Markwardt}, C.~B., {Marshall}, F.~E., {et~al.}
  2003{\natexlab{a}}, \iaucirc, 8064

\bibitem[{{Corbet}(1984)}]{Corbet84}
{Corbet}, R.~H.~D. 1984, \aap, 141, 91

\bibitem[{{Corbet} {et~al.}(2003{\natexlab{b}}){Corbet}, {Markwardt},
  {Marshall}, {Coe}, {Edge}, \& {Laycock}}]{ATel163}
{Corbet}, R.~H.~D., {Markwardt}, C.~B., {Marshall}, F.~E., {et~al.}
  2003{\natexlab{b}}, The Astronomer's Telegram, 163

\bibitem[{{Cutri} {et~al.}(2003){Cutri}, {Skrutskie}, {van Dyk}, {Beichman},
  {Carpenter}, {Chester}, {Cambresy}, {Evans}, {Fowler}, {Gizis}, {Howard},
  {Huchra}, {Jarrett}, {Kopan}, {Kirkpatrick}, {Light}, {Marsh}, {McCallon},
  {Schneider}, {Stiening}, {Sykes}, {Weinberg}, {Wheaton}, {Wheelock}, \&
  {Zacarias}}]{Cutri03}
{Cutri}, R.~M., {Skrutskie}, M.~F., {van Dyk}, S., {et~al.} 2003, VizieR Online
  Data Catalog, 2246

\bibitem[{{Edge} {et~al.}(2004){Edge}, {Coe}, {Galache}, {McBride}, {Corbet},
  {Markwardt}, \& {Laycock}}]{Edge04}
{Edge}, W.~R.~T., {Coe}, M.~J., {Galache}, J.~L., {et~al.} 2004, \mnras, 353,
  1286

\bibitem[{{Evans} {et~al.}(2004){Evans}, {Howarth}, {Irwin}, {Burnley}, \&
  {Harries}}]{Evans04}
{Evans}, C.~J., {Howarth}, I.~D., {Irwin}, M.~J., {Burnley}, A.~W., \&
  {Harries}, T.~J. 2004, \mnras, 353, 601

\bibitem[{{Evans} {et~al.}(2016{\natexlab{a}}){Evans}, {Kennea}, \&
  {Coe}}]{EvansATEL9197}
{Evans}, P.~A., {Kennea}, J.~A., \& {Coe}, M.~J. 2016{\natexlab{a}}, The
  Astronomer's Telegram, 9197

\bibitem[{{Evans} {et~al.}(2007){Evans}, {Beardmore}, {Page}, {Tyler},
  {Osborne}, {Goad}, {O'Brien}, {Vetere}, {Racusin}, {Morris}, {Burrows},
  {Capalbi}, {Perri}, {Gehrels}, \& {Romano}}]{Evans07}
{Evans}, P.~A., {Beardmore}, A.~P., {Page}, K.~L., {et~al.} 2007, \aap, 469,
  379

\bibitem[{{Evans} {et~al.}(2009){Evans}, {Beardmore}, {Page}, {Osborne},
  {O'Brien}, {Willingale}, {Starling}, {Burrows}, {Godet}, {Vetere}, {Racusin},
  {Goad}, {Wiersema}, {Angelini}, {Capalbi}, {Chincarini}, {Gehrels}, {Kennea},
  {Margutti}, {Morris}, {Mountford}, {Pagani}, {Perri}, {Romano}, \&
  {Tanvir}}]{Evans09}
---. 2009, \mnras, 397, 1177

\bibitem[{{Evans} {et~al.}(2014){Evans}, {Osborne}, {Beardmore}, {Page},
  {Willingale}, {Mountford}, {Pagani}, {Burrows}, {Kennea}, {Perri},
  {Tagliaferri}, \& {Gehrels}}]{Evans14}
{Evans}, P.~A., {Osborne}, J.~P., {Beardmore}, A.~P., {et~al.} 2014, \apjs,
  210, 8

\bibitem[{{Evans} {et~al.}(2016{\natexlab{b}}){Evans}, {Osborne}, {Kennea},
  {Campana}, {O'Brien}, {Tanvir}, {Racusin}, {Burrows}, {Cenko}, \&
  {Gehrels}}]{Evans16}
{Evans}, P.~A., {Osborne}, J.~P., {Kennea}, J.~A., {et~al.} 2016{\natexlab{b}},
  \mnras, 455, 1522

\bibitem[{{Galache} {et~al.}(2008){Galache}, {Corbet}, {Coe}, {Laycock},
  {Schurch}, {Markwardt}, {Marshall}, \& {Lochner}}]{Galache08}
{Galache}, J.~L., {Corbet}, R.~H.~D., {Coe}, M.~J., {et~al.} 2008, \apjs, 177,
  189

\bibitem[{{Gehrels} {et~al.}(2004){Gehrels}, {Chincarini}, {Giommi}, {Mason},
  {Nousek}, {Wells}, {White}, {Barthelmy}, {Burrows}, {Cominsky}, {Hurley},
  {Marshall}, {M{\'e}sz{\'a}ros}, {Roming}, {Angelini}, {Barbier}, {Belloni},
  {Campana}, {Caraveo}, {Chester}, {Citterio}, {Cline}, {Cropper}, {Cummings},
  {Dean}, {Feigelson}, {Fenimore}, {Frail}, {Fruchter}, {Garmire}, {Gendreau},
  {Ghisellini}, {Greiner}, {Hill}, {Hunsberger}, {Krimm}, {Kulkarni}, {Kumar},
  {Lebrun}, {Lloyd-Ronning}, {Markwardt}, {Mattson}, {Mushotzky}, {Norris},
  {Osborne}, {Paczynski}, {Palmer}, {Park}, {Parsons}, {Paul}, {Rees},
  {Reynolds}, {Rhoads}, {Sasseen}, {Schaefer}, {Short}, {Smale}, {Smith},
  {Stella}, {Tagliaferri}, {Takahashi}, {Tashiro}, {Townsley}, {Tueller},
  {Turner}, {Vietri}, {Voges}, {Ward}, {Willingale}, {Zerbi}, \&
  {Zhang}}]{Gehrels04}
{Gehrels}, N., {Chincarini}, G., {Giommi}, P., {et~al.} 2004, \apj, 611, 1005

\bibitem[{{Goad} {et~al.}(2007){Goad}, {Tyler}, {Beardmore}, {Evans}, {Rosen},
  {Osborne}, {Starling}, {Marshall}, {Yershov}, {Burrows}, {Gehrels}, {Roming},
  {Moretti}, {Capalbi}, {Hill}, {Kennea}, {Koch}, \& {vanden Berk}}]{Goad07}
{Goad}, M.~R., {Tyler}, L.~G., {Beardmore}, A.~P., {et~al.} 2007, \aap, 476,
  1401

\bibitem[{{Gotthelf} {et~al.}(1999){Gotthelf}, {Vasisht}, \&
  {Dotani}}]{Gotthelf99}
{Gotthelf}, E.~V., {Vasisht}, G., \& {Dotani}, T. 1999, \apjl, 522, L49

\bibitem[{{Grimm} {et~al.}(2003){Grimm}, {Gilfanov}, \& {Sunyaev}}]{Grimm03}
{Grimm}, H.-J., {Gilfanov}, M., \& {Sunyaev}, R. 2003, \mnras, 339, 793

\bibitem[{{Haberl} {et~al.}(2000){Haberl}, {Filipovi{\'c}}, {Pietsch}, \&
  {Kahabka}}]{Haberl00}
{Haberl}, F., {Filipovi{\'c}}, M.~D., {Pietsch}, W., \& {Kahabka}, P. 2000,
  \aaps, 142, 41

\bibitem[{{Haberl} {et~al.}(2007){Haberl}, {Pietsch}, \& {Kahabka}}]{Atel1095}
{Haberl}, F., {Pietsch}, W., \& {Kahabka}, P. 2007, The Astronomer's Telegram,
  1095

\bibitem[{{Haberl} \& {Sturm}(2016)}]{Haberl16}
{Haberl}, F., \& {Sturm}, R. 2016, \aap, 586, A81

\bibitem[{{Harris} \& {Zaritsky}(2004)}]{Harris04}
{Harris}, J., \& {Zaritsky}, D. 2004, \aj, 127, 1531

\bibitem[{{Haschke} {et~al.}(2012){Haschke}, {Grebel}, \& {Duffau}}]{Haschke12}
{Haschke}, R., {Grebel}, E.~K., \& {Duffau}, S. 2012, \aj, 144, 107

\bibitem[{{Hong} {et~al.}(2017){Hong}, {Antoniou}, {Zezas}, {Haberl}, {Sasaki},
  {Drake}, {Plucinsky}, \& {Laycock}}]{Hong17}
{Hong}, J., {Antoniou}, V., {Zezas}, A., {et~al.} 2017, \apj, 847, 26

\bibitem[{{Kahabka} \& {Pietsch}(1996)}]{Kahabka96}
{Kahabka}, P., \& {Pietsch}, W. 1996, \aap, 312, 919

\bibitem[{{Kato} {et~al.}(2007){Kato}, {Nagashima}, {Nagayama}, {Kurita},
  {Koerwer}, {Kawai}, {Yamamuro}, {Zenno}, {Nishiyama}, {Baba}, {Kadowaki},
  {Haba}, {Hatano}, {Shimizu}, {Nishimura}, {Nagata}, {Sato}, {Murai},
  {Kawazu}, {Nakajima}, {Nakaya}, {Kandori}, {Kusakabe}, {Ishihara},
  {Kaneyasu}, {Hashimoto}, {Tamura}, {Tanab{\'e}}, {Ita}, {Matsunaga},
  {Nakada}, {Sugitani}, {Wakamatsu}, {Glass}, {Feast}, {Menzies}, {Whitelock},
  {Fourie}, {Stoffels}, {Evans}, \& {Hasegawa}}]{Kato07}
{Kato}, D., {Nagashima}, C., {Nagayama}, T., {et~al.} 2007, \pasj, 59, 615

\bibitem[{{Kennea} {et~al.}(2016{\natexlab{a}}){Kennea}, {Evans}, \&
  {Coe}}]{KenneaATel9299}
{Kennea}, J.~A., {Evans}, P.~A., \& {Coe}, M.~J. 2016{\natexlab{a}}, The
  Astronomer's Telegram, 9299

\bibitem[{{Kennea} {et~al.}(2017){Kennea}, {Evans}, \& {Coe}}]{KenneaATel10250}
---. 2017, The Astronomer's Telegram, 10250

\bibitem[{{Kennea} {et~al.}(2016{\natexlab{b}}){Kennea}, {Coe}, {Evans},
  {Beardmore}, {Krimm}, {Romano}, {Yamaoka}, {Serino}, \&
  {Negoro}}]{KenneaATel9362}
{Kennea}, J.~A., {Coe}, M.~J., {Evans}, P.~A., {et~al.} 2016{\natexlab{b}}, The
  Astronomer's Telegram, 9362

\bibitem[{{Klus} {et~al.}(2014){Klus}, {Ho}, {Coe}, {Corbet}, \&
  {Townsend}}]{Klus14}
{Klus}, H., {Ho}, W.~C.~G., {Coe}, M.~J., {Corbet}, R.~H.~D., \& {Townsend},
  L.~J. 2014, \mnras, 437, 3863

\bibitem[{{Koliopanos} \& {Vasilopoulos}(2018)}]{Koliopanos18}
{Koliopanos}, F., \& {Vasilopoulos}, G. 2018, \aap, 614, A23

\bibitem[{{Koz{\l}owski} {et~al.}(2011){Koz{\l}owski}, {Kochanek}, \&
  {Udalski}}]{MQS1}
{Koz{\l}owski}, S., {Kochanek}, C.~S., \& {Udalski}, A. 2011, \apjs, 194, 22

\bibitem[{{Koz{\l}owski} {et~al.}(2012){Koz{\l}owski}, {Kochanek}, {Jacyszyn},
  {Udalski}, {Szyma{\'n}ski}, {Poleski}, {Kubiak}, {Soszy{\'n}ski},
  {Pietrzy{\'n}ski}, {Wyrzykowski}, {Ulaczyk}, \& {Pietrukowicz}}]{MQS2}
{Koz{\l}owski}, S., {Kochanek}, C.~S., {Jacyszyn}, A.~M., {et~al.} 2012, \apj,
  746, 27

\bibitem[{{Koz{\l}owski} {et~al.}(2013){Koz{\l}owski}, {Onken}, {Kochanek},
  {Udalski}, {Szyma{\'n}ski}, {Kubiak}, {Pietrzy{\'n}ski}, {Soszy{\'n}ski},
  {Wyrzykowski}, {Ulaczyk}, {Poleski}, {Pietrukowicz}, {Skowron}, {OGLE
  Collaboration}, {Meixner}, \& {Bonanos}}]{MQS3}
{Koz{\l}owski}, S., {Onken}, C.~A., {Kochanek}, C.~S., {et~al.} 2013, \apj,
  775, 92

\bibitem[{{Krimm} {et~al.}(2013){Krimm}, {Holland}, {Corbet}, {Pearlman},
  {Romano}, {Kennea}, {Bloom}, {Barthelmy}, {Baumgartner}, {Cummings},
  {Gehrels}, {Lien}, {Markwardt}, {Palmer}, {Sakamoto}, {Stamatikos}, \&
  {Ukwatta}}]{Krimm13}
{Krimm}, H.~A., {Holland}, S.~T., {Corbet}, R.~H.~D., {et~al.} 2013, \apjs,
  209, 14

\bibitem[{{Li} {et~al.}(1977){Li}, {Jernigan}, \& {Clark}}]{Li77}
{Li}, F., {Jernigan}, G., \& {Clark}, G. 1977, \iaucirc, 3125

\bibitem[{{Li} \& {van den Heuvel}(1997)}]{Li97}
{Li}, X.-D., \& {van den Heuvel}, E.~P.~J. 1997, \aap, 321, L25

\bibitem[{{Lomb}(1976)}]{Lomb76}
{Lomb}, N.~R. 1976, \apss, 39, 447

\bibitem[{{Majid} {et~al.}(2004){Majid}, {Lamb}, \& {Macomb}}]{Majid04}
{Majid}, W.~A., {Lamb}, R.~C., \& {Macomb}, D.~J. 2004, \apj, 609, 133

\bibitem[{{Marshall} {et~al.}(1998){Marshall}, {Lochner}, {Santangelo},
  {Cusumano}, {Israel}, {dal Fiume}, {Orlandini}, {Frontera}, {Parmar}, \&
  {Corbet}}]{MarshallIAUC6818}
{Marshall}, F.~E., {Lochner}, J.~C., {Santangelo}, A., {et~al.} 1998, \iaucirc,
  6818

\bibitem[{{Massey}(2002)}]{Massey02}
{Massey}, P. 2002, \apjs, 141, 81

\bibitem[{{Matsuoka} {et~al.}(2009){Matsuoka}, {Kawasaki}, {Ueno}, {Tomida},
  {Kohama}, {Suzuki}, {Adachi}, {Ishikawa}, {Mihara}, {Sugizaki}, {Isobe},
  {Nakagawa}, {Tsunemi}, {Miyata}, {Kawai}, {Kataoka}, {Morii}, {Yoshida},
  {Negoro}, {Nakajima}, {Ueda}, {Chujo}, {Yamaoka}, {Yamazaki}, {Nakahira},
  {You}, {Ishiwata}, {Miyoshi}, {Eguchi}, {Hiroi}, {Katayama}, \&
  {Ebisawa}}]{Matsuoka09}
{Matsuoka}, M., {Kawasaki}, K., {Ueno}, S., {et~al.} 2009, \pasj, 61, 999

\bibitem[{{McBride} {et~al.}(2017){McBride}, {Gonz{\'a}lez-Gal{\'a}n}, {Bird},
  {Coe}, {Bartlett}, {Dorda}, {Haberl}, {Marco}, {Negueruela}, {Schurch},
  {Sturm}, {Buckley}, \& {Udalski}}]{McBride17}
{McBride}, V.~A., {Gonz{\'a}lez-Gal{\'a}n}, A., {Bird}, A.~J., {et~al.} 2017,
  \mnras, 467, 1526

\bibitem[{{McGowan} {et~al.}(2008){McGowan}, {Coe}, {Schurch}, {McBride},
  {Galache}, {Edge}, {Corbet}, {Laycock}, \& {Buckley}}]{McGowan08}
{McGowan}, K.~E., {Coe}, M.~J., {Schurch}, M.~P.~E., {et~al.} 2008, \mnras,
  383, 330

\bibitem[{{Meegan} {et~al.}(2009){Meegan}, {Lichti}, {Bhat}, {Bissaldi},
  {Briggs}, {Connaughton}, {Diehl}, {Fishman}, {Greiner}, {Hoover}, {van der
  Horst}, {von Kienlin}, {Kippen}, {Kouveliotou}, {McBreen}, {Paciesas},
  {Preece}, {Steinle}, {Wallace}, {Wilson}, \& {Wilson-Hodge}}]{Meegan09}
{Meegan}, C., {Lichti}, G., {Bhat}, P.~N., {et~al.} 2009, \apj, 702, 791

\bibitem[{{Meyssonnier} \& {Azzopardi}(1993)}]{Meyssonnier93}
{Meyssonnier}, N., \& {Azzopardi}, M. 1993, \aaps, 102, 451

\bibitem[{{Negoro} {et~al.}(2016){Negoro}, {Nakajima}, {Kawai}, {Ueno},
  {Tomida}, {Nakahira}, {Ishikawa}, {Sugawara}, {Mihara}, {Sugizaki}, {Serino},
  {Iwakiri}, {Shidatsu}, {Sugimoto}, {Takagi}, {Matsuoka}, {Isobe}, {Sugita},
  {Yoshii}, {Tachibana}, {Ono}, {Fujiwara}, {Harita}, {Muraki}, {Yoshida},
  {Sakamoto}, {Kawakubo}, {Kitaoka}, {Tsunemi}, {Shomura}, {Tanaka},
  {Masumitsu}, {Kawase}, {Ueda}, {Kawamuro}, {Hori}, {Tanimoto}, {Tsuboi},
  {Nakamura}, {Sasaki}, {Yamauchi}, {Furuya}, {Yamaoka}, \&
  {Nakagawa}}]{NegoroATEL9348}
{Negoro}, H., {Nakajima}, M., {Kawai}, N., {et~al.} 2016, The Astronomer's
  Telegram, 9348

\bibitem[{{Nikolajuk} {et~al.}(2015){Nikolajuk}, {Bozzo}, \&
  {Ferrigno}}]{Atel7481}
{Nikolajuk}, M., {Bozzo}, E., \& {Ferrigno}, C. 2015, The Astronomer's
  Telegram, 7481

\bibitem[{{Plucinsky} {et~al.}(2017){Plucinsky}, {Beardmore}, {Foster},
  {Haberl}, {Miller}, {Pollock}, \& {Sembay}}]{2017A&A...597A..35P}
{Plucinsky}, P.~P., {Beardmore}, A.~P., {Foster}, A., {et~al.} 2017, \aap, 597,
  A35

\bibitem[{{Reig}(2011)}]{Reig11}
{Reig}, P. 2011, \apss, 332, 1

\bibitem[{{Roming} {et~al.}(2005){Roming}, {Kennedy}, {Mason}, {Nousek}, {Ahr},
  {Bingham}, {Broos}, {Carter}, {Hancock}, {Huckle}, {Hunsberger}, {Kawakami},
  {Killough}, {Koch}, {McLelland}, {Smith}, {Smith}, {Soto}, {Boyd},
  {Breeveld}, {Holland}, {Ivanushkina}, {Pryzby}, {Still}, \&
  {Stock}}]{Roming05}
{Roming}, P.~W.~A., {Kennedy}, T.~E., {Mason}, K.~O., {et~al.} 2005, \ssr, 120,
  95

\bibitem[{{Rosen} {et~al.}(2016){Rosen}, {Webb}, {Watson}, {Ballet}, {Barret},
  {Braito}, {Carrera}, {Ceballos}, {Coriat}, {Della Ceca}, {Denkinson},
  {Esquej}, {Farrell}, {Freyberg}, {Gris{\'e}}, {Guillout}, {Heil},
  {Koliopanos}, {Law-Green}, {Lamer}, {Lin}, {Martino}, {Michel}, {Motch},
  {Nebot Gomez-Moran}, {Page}, {Page}, {Page}, {Pakull}, {Pye}, {Read},
  {Rodriguez}, {Sakano}, {Saxton}, {Schwope}, {Scott}, {Sturm}, {Traulsen},
  {Yershov}, \& {Zolotukhin}}]{XMM-DR5}
{Rosen}, S.~R., {Webb}, N.~A., {Watson}, M.~G., {et~al.} 2016, \aap, 590, A1

\bibitem[{{Scargle}(1982)}]{Scargle82}
{Scargle}, J.~D. 1982, \apj, 263, 835

\bibitem[{{Schmidtke} \& {Cowley}(2006)}]{Schmidtke06}
{Schmidtke}, P.~C., \& {Cowley}, A.~P. 2006, \aj, 132, 919

\bibitem[{{Schmidtke} {et~al.}(2015){Schmidtke}, {Cowley}, \&
  {Udalski}}]{Atel7498}
{Schmidtke}, P.~C., {Cowley}, A.~P., \& {Udalski}, A. 2015, The Astronomer's
  Telegram, 7498

\bibitem[{{Schurch} {et~al.}(2011){Schurch}, {Coe}, {McBride}, {Townsend},
  {Udalski}, {Haberl}, \& {Corbet}}]{Schurch11}
{Schurch}, M.~P.~E., {Coe}, M.~J., {McBride}, V.~A., {et~al.} 2011, \mnras,
  412, 391

\bibitem[{{Schurch} {et~al.}(2007){Schurch}, {Coe}, {McGowan}, {McBride},
  {Buckley}, {Galache}, {Corbet}, {Still}, {Vaisanen}, {Kniazev}, \&
  {Nordsieck}}]{Schurch07}
{Schurch}, M.~P.~E., {Coe}, M.~J., {McGowan}, K.~E., {et~al.} 2007, \mnras,
  381, 1561

\bibitem[{{Secrest} {et~al.}(2015){Secrest}, {Dudik}, {Dorland}, {Zacharias},
  {Makarov}, {Fey}, {Frouard}, \& {Finch}}]{Secrest15}
{Secrest}, N.~J., {Dudik}, R.~P., {Dorland}, B.~N., {et~al.} 2015, \apjs, 221,
  12

\bibitem[{{Shtykovskiy} \& {Gilfanov}(2005)}]{Shtykovskiy05}
{Shtykovskiy}, P., \& {Gilfanov}, M. 2005, \mnras, 362, 879

\bibitem[{{Sonbas} {et~al.}(2016){Sonbas}, {Lien}, \& {Page}}]{GCN19222}
{Sonbas}, E., {Lien}, A.~Y., \& {Page}, K.~L. 2016, GRB Coordinates Network,
  Circular Service, No.~19222, \#1 (2016), 19222

\bibitem[{{Stella} {et~al.}(1986){Stella}, {White}, \& {Rosner}}]{Stella86}
{Stella}, L., {White}, N.~E., \& {Rosner}, R. 1986, \apj, 308, 669

\bibitem[{{Sturm} {et~al.}(2013){Sturm}, {Haberl}, {Pietsch}, {Ballet},
  {Hatzidimitriou}, {Buckley}, {Coe}, {Ehle}, {Filipovi{\'c}}, {La Palombara},
  \& {Tiengo}}]{Sturm13}
{Sturm}, R., {Haberl}, F., {Pietsch}, W., {et~al.} 2013, \aap, 558, A3

\bibitem[{{Townsend} {et~al.}(2013){Townsend}, {Drave}, {Hill}, {Coe},
  {Corbet}, {Bird}, \& {Schurch}}]{Townsend13}
{Townsend}, L.~J., {Drave}, S.~P., {Hill}, A.~B., {et~al.} 2013, \mnras, 433,
  23

\bibitem[{{Townsend} {et~al.}(2017){Townsend}, {Kennea}, {Coe}, {McBride},
  {Buckley}, {Evans}, \& {Udalski}}]{Townsend2017}
{Townsend}, L.~J., {Kennea}, J.~A., {Coe}, M.~J., {et~al.} 2017, \mnras, 471,
  3878

\bibitem[{{Trowbridge} {et~al.}(2007){Trowbridge}, {Nowak}, \&
  {Wilms}}]{Trowbridge07}
{Trowbridge}, S., {Nowak}, M.~A., \& {Wilms}, J. 2007, \apj, 670, 624

\bibitem[{{Tsygankov} {et~al.}(2017){Tsygankov}, {Doroshenko}, {Lutovinov},
  {Mushtukov}, \& {Poutanen}}]{Tsygankov17}
{Tsygankov}, S.~S., {Doroshenko}, V., {Lutovinov}, A.~A., {Mushtukov}, A.~A.,
  \& {Poutanen}, J. 2017, \aap, 605, A39

\bibitem[{{Udalski} {et~al.}(1997){Udalski}, {Kubiak}, \&
  {Szymanski}}]{Udalski97}
{Udalski}, A., {Kubiak}, M., \& {Szymanski}, M. 1997, \actaa, 47, 319

\bibitem[{{Udalski} {et~al.}(2015){Udalski}, {Szyma{\'n}ski}, \&
  {Szyma{\'n}ski}}]{Udalski15}
{Udalski}, A., {Szyma{\'n}ski}, M.~K., \& {Szyma{\'n}ski}, G. 2015, \actaa, 65,
  1

\bibitem[{{VanderPlas}(2017)}]{VanderPlas17}
{VanderPlas}, J.~T. 2017, ArXiv e-prints, arXiv:1703.09824

\bibitem[{{Vasilopoulos} {et~al.}(2017{\natexlab{a}}){Vasilopoulos}, {Haberl},
  \& {Maggi}}]{Vasilopoulos17b}
{Vasilopoulos}, G., {Haberl}, F., \& {Maggi}, P. 2017{\natexlab{a}}, \mnras,
  470, 1971

\bibitem[{{Vasilopoulos} {et~al.}(2017{\natexlab{b}}){Vasilopoulos}, {Zezas},
  {Antoniou}, \& {Haberl}}]{Vasilopoulos17a}
{Vasilopoulos}, G., {Zezas}, A., {Antoniou}, V., \& {Haberl}, F.
  2017{\natexlab{b}}, \mnras, 470, 4354

\bibitem[{{V{\'e}ron-Cetty} \& {V{\'e}ron}(2010)}]{Veron10}
{V{\'e}ron-Cetty}, M.-P., \& {V{\'e}ron}, P. 2010, \aap, 518, A10

\bibitem[{{Voges} {et~al.}(1999){Voges}, {Aschenbach}, {Boller},
  {Br{\"a}uninger}, {Briel}, {Burkert}, {Dennerl}, {Englhauser}, {Gruber},
  {Haberl}, {Hartner}, {Hasinger}, {K{\"u}rster}, {Pfeffermann}, {Pietsch},
  {Predehl}, {Rosso}, {Schmitt}, {Tr{\"u}mper}, \& {Zimmermann}}]{Voges99}
{Voges}, W., {Aschenbach}, B., {Boller}, T., {et~al.} 1999, \aap, 349, 389

\bibitem[{{Weng} {et~al.}(2017){Weng}, {Ge}, {Zhao}, {Wang}, {Zhang}, {Bian},
  \& {Yuan}}]{Weng17}
{Weng}, S.-S., {Ge}, M.-Y., {Zhao}, H.-H., {et~al.} 2017, \apj, 843, 69

\bibitem[{{Willingale} {et~al.}(2013){Willingale}, {Starling}, {Beardmore},
  {Tanvir}, \& {O'Brien}}]{Willingale13}
{Willingale}, R., {Starling}, R.~L.~C., {Beardmore}, A.~P., {Tanvir}, N.~R., \&
  {O'Brien}, P.~T. 2013, \mnras, 431, 394

\bibitem[{{Yang} {et~al.}(2017){Yang}, {Laycock}, {Christodoulou}, {Fingerman},
  {Coe}, \& {Drake}}]{Yang17}
{Yang}, J., {Laycock}, S.~G.~T., {Christodoulou}, D.~M., {et~al.} 2017, \apj,
  839, 119

\end{thebibliography}

\startlongtable
\begin{deluxetable*}{cccccccccc}
\tabletypesize{\scriptsize}
\tablewidth{8in}
\tablecolumns{10}
\tablecaption{\label{tab:cattable}Catalog of X-ray sources detected in the first year of S-CUBED observations. Only sources that have been flagged as `Good' quality detections are listed in this table. \label{cattable}}
\tablehead{
\colhead{SC\#} & 
\colhead{Name} & 
\colhead{RA/Dec} & 
\colhead{Err} & 
\colhead{Det.} & 
\colhead{$f_\mathrm{mean}$ (0.5-10~keV)}  & 
\colhead{$\Gamma$} & 
\colhead{Catalog Name} & 
\colhead{Cl.} & 
\colhead{Type [Ref]}
\\
\colhead{} & 
\colhead{(1SCUBEDX J...)} & 
\colhead{(J2000)} & 
\colhead{($''$)} & 
\colhead{(\%)} & 
\colhead{(erg/s/cm$^2$)}   & 
\colhead{} & 
\colhead{}      & 
\colhead{} & 
\colhead{}
}
\startdata
SC1   & J011705.2-732635     & $01^{h} 17^{m} 05\fs18\ {-73}^\circ 26' 35\farcs8$ & $2.2^*$ & 100.0 & $6.3^{+0.1}_{-0.1} \times 10^{-10}$ & $1.15^{+0.05}_{-0.03}$    & SMC X-1                  & K     & HMXB  [2]       \\
SC2   & J010401.3-720155     & $01^{h} 04^{m} 01\fs29\ {-72}^\circ 01' 55\farcs7$ & $3.8$   & 100.0 & $3.3^{+0.1}_{-0.1} \times 10^{-11}$ & $6.94^{+0.32}_{-0.29}$    & 1E 0102.2-7219           & K     & SNR             \\
SC3   & J004854.9-734945     & $00^{h} 48^{m} 54\fs93\ {-73}^\circ 49' 45\farcs6$ & $3.8$   & 59.1  & $3.2^{+0.2}_{-0.2} \times 10^{-12}$ & $0.94^{+0.20}_{-0.15}$    & SXP 15.6                 & K     & BeXRB  [1, 2]   \\
SC4   & J003817.7-730926     & $00^{h} 38^{m} 17\fs67\ {-73}^\circ 09' 26\farcs6$ & $7.1$   & 19.5  & $9.3^{+1.2}_{-1.1} \times 10^{-13}$ & $1.57^{+0.68}_{-0.27}$    & 1RXS J003823.0-730931    & K     & -               \\
SC5   & J004910.3-724939     & $00^{h} 49^{m} 10\fs29\ {-72}^\circ 49' 39\farcs2$ & $4.1$   & 78.0  & $7.1^{+0.4}_{-0.4} \times 10^{-12}$ & $0.70^{+0.30}_{-0.30}$    & SXP 18.3                 & K     & BeXRB  [1]      \\
SC6   & J005056.4-721333     & $00^{h} 50^{m} 56\fs39\ {-72}^\circ 13' 33\farcs4$ & $4.6$   & 43.2  & $2.8^{+0.2}_{-0.2} \times 10^{-12}$ & $0.85^{+0.23}_{-0.18}$    & SXP 91.1                 & K     & BeXRB  [1, 2]   \\
SC7   & J003850.6-731051     & $00^{h} 38^{m} 50\fs65\ {-73}^\circ 10' 51\farcs1$ & $6.5$   & 16.7  & $4.3^{+0.7}_{-0.7} \times 10^{-13}$ & $2.41^{+1.19}_{-0.74}$    & OGLE 003850.79-731053.1  & K     & AGN  [9, 10]    \\
SC8   & J003235.5-730650     & $00^{h} 32^{m} 35\fs46\ {-73}^\circ 06' 50\farcs9$ & $2.3^*$ & 71.4  & $1.4^{+0.1}_{-0.1} \times 10^{-12}$ & $1.53^{+0.20}_{-0.10}$    & -                        & U     & AGN             \\
\enddata
\tablecomments{References source type: [1] \cite{CoeKirk15}, [2] \cite{Haberl16}, [3] \cite{Azzopardi79}, [4] \cite{Meyssonnier93}, [5] \cite{Massey02}, [6] \cite{MQS1}, [7] \cite{MQS2}, [8] \cite{MQS3}, [9] \cite{Veron10}, [10] \cite{Cioni13}, [11] \cite{McGowan08}, [12] \cite{Sturm13} }
\end{deluxetable*}

\end{document}